\documentclass[prd,onecolumn,amsmath,amssymb,floatfix,nofootinbib]{revtex4}

\usepackage{amssymb}
\usepackage{amsmath}

\usepackage{graphicx}
\usepackage{graphics}
\usepackage{dcolumn}
\usepackage{color}
\usepackage{rotate}
\usepackage{fancyhdr}

\def\be{\begin{equation}}
\def\ee{\end{equation}}
\def\bea{\begin{eqnarray}}
\def\eea{\end{eqnarray}}
\def\ba{\begin{eqnarray}}
\def\ea{\end{eqnarray}}
\def\nn{\nonumber}
\def\vs{\nonumber\\ }

\def\VEV#1{\left\langle #1 \right\rangle}

\newcommand{\hatk}{{\mathbf{\hat k}}}

\newcommand{\hattheta}{{\mathbf{\hat \theta}}}
\newcommand{\hatphi}{{\mathbf{\hat \phi}}}
\newcommand{\hatz}{{\mathbf{\hat z}}}
\newcommand{\hatn}{{\mathbf{\hat n}}}

\newcommand{\hatq}{{\mathbf{\hat q}}}
\newcommand{\hatx}{{\mathbf{\hat x}}}

\newcommand{\bfx}{\mathbf{x}}
\newcommand{\bfk}{\mathbf{k}}

\definecolor{darkred}{rgb}{.743,0,0}

\def\bm#1{\mathbf{#1}}

\def\qhat{\hat{\bm{q}}}

\def\lm{{(lm)}}
\def\JM{{(JM)}}
\def\mbar{{\bar{m}}}

\def\wigner#1#2#3#4#5#6{ \left( \begin{array}{ccc} #1 & #3 & #5
\\ #2 & #4 & #6 \\ \end{array} \right)}
\def\wignersj#1#2#3#4#5#6{ \left\{ \begin{array}{ccc} #1 & #3 & #5
\\ #2 & #4 & #6 \\ \end{array} \right\}}

\begin{document}

\title{Total Angular Momentum Waves for Scalar, Vector, and
Tensor Fields}

\author{Liang Dai, Marc Kamionkowski, and Donghui Jeong}
\affiliation{$^1$Department of Physics and Astronomy, Johns
     Hopkins University, 3400 N.\ Charles St., Baltimore, MD 21218}

\date{\today}

\begin{abstract}
Most calculations in cosmological perturbation theory, including
those dealing with the inflationary generation of perturbations,
their time evolution, and their observational consequences,
decompose those perturbations into plane waves (Fourier modes).
However, for some calculations, particularly those involving
observations performed on a spherical sky, a decomposition into
waves of fixed total angular momentum (TAM) may
be more appropriate.  Here we introduce TAM waves---solutions of
fixed total angular momentum to the Helmholtz equation ---for 
three-dimensional scalar, vector, and tensor fields.  The vector
TAM waves of given total angular momentum can be decomposed
further into a set of three basis functions of
fixed orbital angular momentum (OAM), a set of fixed
helicity, or a basis consisting of a longitudinal ($L$) and two
transverse ($E$ and $B$) TAM waves.  The symmetric traceless
rank-2 tensor TAM waves can be similarly decomposed into a basis of 
fixed OAM or fixed helicity, or a basis that consists of a
longitudinal ($L$), two vector ($VE$ and $VB$, of opposite
parity), and  two tensor ($TE$ and $TB$, of opposite parity)
waves. We show how all of the vector and tensor TAM waves can be
obtained by applying derivative operators to scalar TAM waves.
This operator approach then allows one to decompose a vector
field into three covariant scalar fields for the $L$, $E$, and
$B$ components and symmetric-traceless-tensor fields into five
covariant scalar fields for the $L$, $VE$, $VB$, $TE$, and $TB$
components.  We provide projections of the vector and tensor TAM
waves onto vector and tensor spherical harmonics.  We provide
calculational detail to facilitate the assimilation of this
formalism into cosmological calculations.
As an example, we calculate the power spectra of the deflection
angle for gravitational lensing by density perturbations and by
gravitational waves.  We comment on an alternative approach to CMB
fluctuations based on TAM waves.  An accompanying paper will
work out three-point functions in terms of TAM waves and their
relation to the usual Fourier-space bispectra.  Our work may
have applications elsewhere in field theory and in general
relativity.
\end{abstract}
\pacs{}

\maketitle

\tableofcontents

\section{Introduction}

Much of modern cosmology involves the study of the origin and
evolution of scalar, vector, and tensor fields.  Examples of
scalar fields include the inflaton \cite{inflation} and the
quintessence field \cite{quintessence}.
Perturbations in the inflaton are considered as seeds for
primordial perturbations to the curvature,
and there is active investigation of the effects of
quintessence perturbations on the evolution of density
perturbations at late times.  Magnetic fields provide an example
of vector fields in cosmology
\cite{Turner:1987bw,Caldwell:2011ra}, and vector fields have appeared
elsewhere as well \cite{vectors}.  The most general perturbation
to the spacetime metric involves a rank-2 tensor field
\cite{Bertschinger:1993xt}, the six components of which, as is
well known, can be decomposed into a trace, a longitudinal mode,
two vector modes, and two transverse-traceless modes, the latter
of which propagate in general relativity as
gravitational waves.  A stochastic background of cosmological
gravitational waves may have been produced during inflation
\cite{Rubakov:1982df} and are now being actively sought through
the curl component ($B$ mode) they induce in the anisotropy of
the cosmic microwave background (CMB) polarization
\cite{Kamionkowski:1996ks}.

The vast majority of the literature on such fields and
perturbations proceeds by decomposing the perturbations into
Fourier modes (or plane waves) $e^{i\bfk\cdot\bfx}$, each of
which then evolves independently to first order in perturbation
theory.  A primordial random field, such as that produced by
inflation, is then assembled by adding all such plane waves with
the Fourier amplitude for each wavevector $\bfk$ selected from a
Gaussian distribution with a variance given by the power
spectrum $P(k)$.

However, observations of the Universe are performed on a
spherical sky. It may thus be advantageous, in some
cases, to consider decomposition of the scalar/vector/tensor
fields under consideration into a basis that reflects the
rotational symmetry of the spherical sky.  With this
motivation in mind, we introduce here total-angular-momentum
(TAM) waves for scalar, vector, and tensor fields.  We provide a complete
orthonormal set of basis functions for scalar, vector, and
tensor fields on three-dimensional Euclidean space, of fixed
orbital angular momentum.  
These basis functions are eigenfunctions of the Laplacian operator.
There are three vector TAM waves for each total
angular momentum, and we decompose these three into a
basis of fixed orbital angular momentum (OAM), a basis of fixed
helicity, and a basis that separates the longitudinal ($L$) and two
transverse modes, $E$ and $B$, of opposite parity.  There are
similarly five sets of TAM basis functions for traceless rank-2
symmetric tensors, and we provide similar decompositions into
three sets of bases: an OAM basis, a
helicity basis, and a basis that decomposes into a longitudinal
($L$) TAM wave, two vector waves, $VE$ and $VB$, of opposite
parity, and two transverse waves, $TE$ and $TB$, of opposite
parity.  A random field can then be assembled by adding all such
TAM waves, with the amplitude for each TAM wave of
wavenumber $k$ selected from a Gaussian distribution with
variance $P(k)$, as will be detailed below.

The $L/E/B$ basis for vectors, and the $L/VE/VB/TE/TB$
basis for traceless symmetric rank-2 tensors, are first derived from the OAM
basis.  The helicity bases are then simply related to these
$L/E/B$ and $L/VE/VB/TE/TB$ bases.  We then present an
alternative derivation of these TAM-wave bases by introducing
sets of vector and traceless-tensor differential operators that,
when applied to the scalar TAM waves, yield the  $L/E/B$ and
$L/VE/VB/TE/TB$ TAM waves.  This operator approach then allows
one to write an arbitrary vector field $V_a(\bfx)$ in terms of
three covariant scalar functions $V^L(\bfx)$, $V^E(\bfx)$, and
$V^B(\bfx)$ and an arbitrary tensor field $h_{ab}(\bfx)$ in
terms of five covariant scalar functions $h^L(\bfx)$,
$h^{VE}(\bfx)$, $h^{VB}(\bfx)$, $h^{TE}(\bfx)$, $h^{TB}(\bfx)$.
The operator approach also allows one to obtain sets of vector
and tensor basis functions from any other set of scalar basis
functions.

We also provide the projections of the TAM vector and tensor
waves onto vector and tensor spherical harmonics.  This is
equivalent, as will be seen below, to providing the $\hatn$,
$\hattheta$, and $\hatphi$ components of the TAM vector and
tensor waves.  The utility of TAM waves, as well as these
projections, is illustrated with a re-derivation of the power
spectra for weak lensing by density perturbations and
gravitational waves.

The bases we provide here for three-dimensional fields are to
be contrasted with earlier work \cite{Kamionkowski:1996ks},
developed for CMB polarization and weak-lensing shear, on
bases for tensor fields on the two-sphere $\mathbb{S}^2$, and with bases
for vector fields (the weak-lensing deflection angle)
\cite{Hu:2000ee}; we will show below, though, how the
$\hattheta$-$\hatphi$ components of the three-dimensional
TAM waves map onto the familiar vector/tensor spherical
harmonics. 
Ref.~\cite{Tomita:1982ew,Bucher:1997xs,Challinor:1999xz}
present the E/B modes of three-dimensional vector and tensor harmonics 
in open and closed Friedmann-Robertson-Walker space.
The TAM-wave basis for scalar fields has
already been employed in cosmology
\cite{Fisher:1994cm,Heavens:1994iq,Rassat:2011aa,Leistedt:2011mk,Abramo:2010za}
(sometimes referred to as a ``spherical-wave'' or
``Fourier-Bessel'' expansion).  The vector TAM waves are
familiar from electromagnetic theory
\cite{Jackson,Varshalovich}.  Some steps along these lines for
tensor fields were taken in Ref.~\cite{Hu:1997hp}, although they
retained plane waves for the spatial dependence.  There are some
resemblances to Ref.~\cite{DiDio:2012bu}, who were considering
classical cosmological tests.
Refs.~\cite{Heavens:2003jx,Castro:2005bg} have taken significant
steps in the direction we pursue here for the description of
weak lensing by density perturbations, and there are some
analogues to this work in the gravitational-wave literature
(see, e.g., Ref.~\cite{Thorne:1980ru}).

Below we begin in Section \ref{sec:notation} with a brief
discussion of our notation.  Section \ref{sec:scalarfields}
presents TAM waves for scalar fields beginning, by way of
introduction, with a review in Section
\ref{sec:scalarplanewaves} of the Fourier expansion of scalar
fields.  Section \ref{sec:vectorfields} discusses vector fields,
beginning in \ref{sec:vectorplanewaves} with plane waves and
moving on in \ref{sec:vectortamwaves} to TAM waves with vector
fields.  There the TAM waves of fixed total angular momentum are
decomposed into OAM, $L/E/B$, and helicity
bases.  We introduce in Section \ref{sec:vectoroperators} a set
of derivative operators that, when applied to the scalar TAM
waves, provide TAM vector waves in the $L/E/B$ basis.  We also
show here how this operator approach can be used to find scalar
functions associated with the $L$, $E$, and $B$ components.
The rest of Section \ref{sec:vectorfields} discusses the
projection of the  TAM vector waves onto vector spherical
harmonics (\ref{sec:vectorprojection}), results that are useful,
e.g., for observational quantities like the lensing deflection
field that are represented as vectors on the two-sphere; the
transformation between vector plane waves and vector TAM waves
(\ref{sec:vectortransforms}); and the expansion of vector fields
in terms of TAM waves and the relation between the TAM-wave
power spectra and the more familiar plane-wave power spectra
(\ref{sec:vectorexpansion}).  Section \ref{sec:tensorfields}
provides a discussion of tensor TAM waves with an organization
that parallels precisely that for vector waves in Section
\ref{sec:vectorfields}.  Section \ref{sec:lensing} presents, as
an example of the utility of the TAM-wave formalism, a calculation of
the power spectra for the deflection angle from gravitational
lensing by density (scalar) perturbations and gravitational
waves (transverse-traceless tensor perturbations).  Section
\ref{sec:cmb} discusses the prospects for writing the Boltzmann
equations for the evolution of CMB fluctuations using the
TAM-wave formalism.  Section \ref{sec:conclusions} provides
closing remarks.  Appendixes \ref{appendix:vectordivergence} and
\ref{appendix:tensordivergence} provide calculational details,
and Appendix \ref{appendix:irreducible} provides a proof that
the functions obtained by the action of irreducible-tensor operators
on TAM waves are TAM waves of the same total-angular-momentum
quantum numbers $JM$.

\section{Notation}
\label{sec:notation}

Throughout this paper we use the symbol $\Psi^k(\bfx)$ to denote
solutions to the Helmholtz equation,
\be
(\nabla^2+k^2) \Psi^k(\bfx)=0,
\label{def:Helmholtz}
\ee
where $\bfx$ is the spatial position, and $k$ is the magnitude of the
wavevector.  In order to reduce clutter, we will often suppress
the $k$ superscript.  We will be dealing with solutions
$\Psi^k_{(JM)}(\bfx)$ that are eigenstates of total angular momentum and
its $z$ component labeled by eigenvalues $J$ and $M$,
respectively.  We will also obtain scalar, vector, and tensor
solutions to the Helmholtz equation, and we will denote those
solutions (actually, the components of those solutions) by
$\Psi_{(JM)}(\bfx)$, $\Psi_{(JM)a}(\bfx)$, and $\Psi_{(JM)ab}(\bfx)$ 
(where we
have suppressed the $k$ superscript, as we will do frequently
throughout), respectively.  The number of indices in the subscript, outside
the parentheses, indicates whether the quantity is a scalar,
vector, or tensor.  As we will see, the vector and tensor
eigenfunctions of fixed $JM$ can be decomposed into states of
fixed orbital angular momentum, fixed helicity, or a
longitudinal/transverse decomposition.  These will be labeled by
a superscript.  For example, the vector eigenstate of total
angular momentum $JM$ for wavevector $k$ with orbital angular
momentum $l$ will be $\Psi^{l,k}_{(JM)a}(\bfx)$, and the vector
TAM waves in the transverse/longitudinal basis will be referred
to by $\Psi^{\alpha,k}_{(JM)a}(\bfx)$, for $\alpha=L,E,B$, and in the
helicity basis by  $\Psi^{\lambda,k}_{(JM)a}(\bfx)$, for
$\lambda=0,\pm1$.  Again, the $k$
superscript will often be suppressed.  We often refer to $V_a$
as a ``vector,'' although strictly speaking, it is the dual
vector associated with the vector $V^a$; there should never be
any confusion given that the dual vector has a lowered index
and the vector a raised index.  The indices are raised
and lowered with a metric $g_{ab}$, and the antisymmetric tensor
is $\epsilon_{abc}$.  Since we are dealing with
flat three-dimensional space, the metric may be taken to be a
Kronecker delta with Cartesian coordinates, in which case the
raising and lowering of indices is trivial.  However, we will at
times work in spherical coordinates $r,\theta,\phi$ in which
case $g_{ab}$ is not trivial. In some places we will deal with
functions on the two-sphere $\mathbb{S}^2$, and in these cases we denote
the metric and antisymmetric tensor for the two-sphere by
$g_{AB}$ and $\epsilon_{AB}$, respectively, with capital indices.

We will also, by way of introduction, deal with plane-wave
solutions to the Helmholtz equation.  We will label the scalar,
vector, and tensor solutions by $\Psi^{\bfk}(\bfx)$,
$\Psi^{\bfk}_a(\bfx)$, and $\Psi^{\bfk}_{ab}(\bfx)$,
respectively.  An additional superscript will denote the
decomposition into OAM, helicity, or
longitudinal/transverse eigenstates.\footnote{Note that there is
no such superscript for scalar waves, as the OAM and TAM waves
coincide for scalar fields, since they have no spin.}  For
reference, we list in Table I the symbols used in this paper.

\begin{table}[htbp]
\begin{center}
\begin{tabular}{|c|c|c|}
\hline
$\bfx={x,y,z}$, $r=|\bfx|$, and $\hatn$ &  a point in ${\mathbb
R}^3$, its norm, and a unit vector in its direction & \\ \hline
$\bfk$, $k_a$, and $k=|\bfk|$ &  Fourier wavevector, its
components, and its magnitude & \\ \hline
$\hatk$, $\hattheta$, etc. and $\hat k_a$, $\hat\theta_a$,
etc. & unit vectors in the $\bfk$, $\theta$,
etc.\ directions and their components & \\ \hline
$\nabla$, $\nabla_a$ & covariant derivative wrt ${\bfx}$ and its
components  & \\ \hline
$*$ and $\dagger$ & as superscripts represent complex
conjugation and hermitian conjugate & \\ \hline
$\VEV{X}$ & average over all realizations of random variable $X$
& (\ref{def:pk_scalar}) \\ \hline
$\delta_D(k-k')$ & the one-dimensional Dirac delta function &
(\ref{def:norm_jl_Ylm}) \\ \hline
$\delta_D(\bfk-\bfk')$ & the three-dimensional Dirac delta function
& (\ref{def:Phibfk}) \\ \hline
$a,b,c,\ldots$ &  three-dimensional tensor indices & \\ \hline
$A,B,C,\ldots$ &  two-dimensional tensor indices & \\ \hline
$\hat{\varepsilon}_a$ & polarization vector & \\ \hline
$\delta_{ij}$ & Kronecker delta & \\ \hline
$g_{ab}$ and $\epsilon_{abc}$ & metric and
antisymmetric tensor in ${\mathbb R}^3$ & \\ \hline
$g_{AB}$ and $\epsilon_{AB}$ & metric an antisymmetric tensor on
${\mathbb S}^2$ &
(\ref{eqn:EBtensor}) \\ \hline
$\varepsilon_{ab}^s$ & polarization tensors  &
(\ref{eqn:SVTdecomposition}) \\ \hline
$\hat \varepsilon_{ab}(\bfk)$ & polarization tensor for tensor
plane wave & (\ref{eqn:tensorplanewaves}) \\ \hline
$\tilde h_{ab}(\bfk)$ & Fourier transform of $h_{ab}(\bfx)$ &
(\ref{eqn:SVTdecomposition}) \\ \hline
$h_s(\bfk)$ & amplitudes for tensor plane-wave components &
(\ref{eqn:SVTdecomposition}) \\ \hline
${\bf L}$ and $L_a$ &  orbital-angular-momentum operator and its components
& \\ \hline
${\bf S}$ &  spin operator & \\ \hline
${\bf J}$ and $J_a$ &  total-angular-momentum operator
and its components & \\ \hline
$J$ and \ $M$ & Quantum numbers for total
angular momentum and its $z$ component & \\ \hline
$l$ and $m$ & Quantum numbers for orbital
angular momentum and its $z$ component & \\ \hline
$\wigner{l_1}{m_1}{l_2}{m_2}{l_3}{m_3}$ & Wigner-$3j$ symbol
& (\ref{eq:Wigner3j}) \\ \hline
$\wignersj{l_1}{m_1}{l_2}{m_2}{l_3}{m_3}$ & Wigner-$6j$ symbol
& (\ref{eq:Wigner6j}) \\ \hline
$\VEV{l_1 m_1 l_2 m_2| JM}$ & Clebsch-Gordan coefficient
& (\ref{eqn:CBexpansion})\\ \hline
$V_a(\bfx)$ & vector field &
(\ref{eqn:vectorexpansion})\\ \hline
$h_{ab}(\bfx)$ & tensor field &
(\ref{eqn:SVTdecomposition})\\ \hline
$\tilde V^\alpha(\bfk)$ & Fourier coefficients for vector field &
(\ref{eqn:vectorexpansion})\\ \hline
$V^\alpha(\bfx)$ & scalar fields for $\alpha=L,E,B$
components of vector field &
(\ref{eqn:scalarfunctions})\\ \hline
$h^\alpha(\bfx)$ & scalar fields for $\alpha=L,VE,VB,TE,TB$
components of traceless tensor field &
(\ref{eqn:tensorscalarfunctions})\\ \hline
$P(k)$ & power spectrum for density perturbations &
(\ref{def:pk_scalar})\\ \hline
$P_L(k)$,$P_T(k)$ & power spectra for longitudinal and
transverse modes of vector field &
(\ref{def:pk_vector}) \\\hline
$P_+(k)$,$P_-(k)$ & power spectra for left- and right-circularly
polarized vector fields &
(\ref{def:vectorPolarization}) \\\hline
$P_h(k)$ & power spectrum for transverse-traceless mode of tensor field
& (\ref{def:pk_tensor})\\ \hline
$\Psi^k(\bfx)$ & Solutions to the Helmholtz equation for
wavenumber $k$ & (\ref{def:Helmholtz}) \\ \hline
$\phi(\bfx)$ &  scalar functions & (\ref{def:phixphik}) \\ \hline
$\tilde\phi(\bfk)$ & the Fourier transform of $\phi(\bfx)$ &
(\ref{def:phixphik}) \\ \hline
$\phi_{(lm)}(k)$ & TAM-wave transform of $\phi(\bfx)$
& (\ref{eqn:philmdef})\\ \hline
$\xi(\bfx)$ &  scalar for longitudinal component of $h_{ab}$ &
(\ref{eqn:SVTdecomposition}) \\ \hline
$w_a(\bfx)$ &  transverse-vector field for vector component of $h_{ab}$ &
(\ref{eqn:SVTdecomposition}) \\ \hline
$h_{ab}^{TT}(\bfx)$ &  transverse-traceless part of $h_{ab}$ &
(\ref{eqn:SVTdecomposition}) \\ \hline
$\Psi^{\bfk}(\bfx)$ & scalar plane-wave mode & (\ref{def:Phibfk}) \\ \hline
$\Psi^{\alpha,\bfk}_a(\bfx)$ &  vector plane-wave mode for
polarization $\alpha=L,1,2$ &(\ref{eqn:vectorFourier}) \\ \hline
$\Psi^{\lambda,\bfk}_a(\bfx)$ &  circularly-polarized vector
plane-wave mode for helicity $\lambda=\pm1$ & (\ref{def:vectorPolarization})\\ \hline
$j_l(x)$ and $n_l(x)$ & spherical Bessel functions of the first
and second kind
& (\ref{def:norm_jl_Ylm}) \\ \hline
$Y_{(lm)}(\hatn)$ & scalar spherical harmonic & (\ref{def:norm_jl_Ylm})\\ \hline
$e_{a}^{\mbar}$ & spherical basis for vector & (\ref{def:spherical_basis_vector}) \\ \hline
$t_{ab}^{\tilde{m}}$ & spherical basis for tensor & (\ref{def:spherical_basis_tensor}) \\ \hline
$Y_{(JM)a}^l(\hatn)$ & vector spherical harmonic of OAM $l$, for
$l=J-1,J,J+1$ & (\ref{eqn:CBexpansion})\\ \hline
$Y_{(JM)a}^\alpha(\hatn)$ & vector spherical harmonic in the
longitudinal/transverse basis for $\alpha=L,E,B$ & (\ref{eqn:vectorYlms})\\ \hline
$Y_{(JM)a}^\lambda(\hatn)$ & vector spherical harmonic of helicity
$\lambda=0,\pm1$ & (\ref{def:vector_helicity_Ylm})\\ \hline
$Y_{(JM)ab}^l(\hatn)$ & tensor spherical harmonic of OAM $l$, for
$l=J-2,\ldots,J+2$ & (\ref{def:Psiabl})\\ \hline
$Y_{(JM)ab}^\alpha(\hatn)$ & tensor spherical harmonic for
$\alpha=L,VE,VB,TE,TB$ & (\ref{def:Ylmabl})\\ \hline
$Y_{(JM)ab}^\lambda(\hatn)$ & tensor spherical harmonic of helicity
$\lambda=0,\pm1,\pm2$ &(\ref{def:Ylmablambda}) \\ \hline 
\end{tabular}
\end{center}
\end{table}

\begin{table}[htbp]
\label{tab:symbols}
\begin{center}
\begin{tabular}{|c|c|c|}
\hline
$\Psi_{(lm)}(\bfx)$ & scalar TAM wave & (\ref{def:Psilmk})\\ \hline
$\Psi_{(JM)a}^l(\bfx)$ & vector TAM wave of OAM $l$, for
$l=J-1,J,J+1$ & (\ref{eqn:CBexpansion})\\ \hline
$\Psi_{(JM)a}^\alpha(\bfx)$ & vector TAM wave in the
longitudinal/transverse basis for $\alpha=L,E,B$ & (\ref{eqn:vectorTAMs})\\ \hline
$\Psi_{(JM)a}^\lambda(\bfx)$ & vector TAM wave of helicity
$\lambda=0,\pm1$ & (\ref{def:Psialambda})\\ \hline
$V^l_{(JM)}$, $V^\alpha_{(JM)}$, $V^\lambda_{(JM)}$ & vector
TAM-wave coefficients & (\ref{eqn:vectorcoeffs}) \\ \hline
$\Psi_{(JM)ab}^l(\bfx)$ & tensor TAM wave of OAM $l$, for
$l=J-2,\ldots,J+2$ & (\ref{def:Psiabl})\\ \hline 
$\Psi_{(JM)ab}^\alpha(\bfx)$ & tensor TAM wave for
$\alpha=L,VE,VB,TE,TB$ & (\ref{def:Psiabalpha})\\ \hline
$\Psi_{(JM)ab}^\lambda(\bfx)$ & tensor TAM wave of helicity
$\lambda=0,\pm1,\pm2$ & (\ref{def:Psiablambda}) \\ \hline
$h_{(JM)}^l$, $h_{(JM)}^\alpha$, $h_{(JM)}^\lambda$ & tensor TAM-wave
expansion coefficients  & (\ref{eqn:tensorexpansioncoeffs}) \\ \hline
$A_{(JM)}^l$, $A_{(JM)}^\alpha$, $A_{(JM)}^\lambda$, 
 & plane-wave expansion coefficients for vectors
& (\ref{eqn:vectorplanewaves}) \\ \hline
$B_{(JM)}^l$, $B_{(JM)}^\alpha$, $B_{(JM)}^\lambda$, 
 & plane-wave expansion coefficients for tensors
& (\ref{eqn:tensorplanewaves}) \\ \hline
$D_a$, $M_a$, and $K_a$ & differential operators &
(\ref{eqn:operatordefs}) \\ \hline
$N_a$, $M_{\perp a}$, and $K_a$ & operators normal ($N_a$) tangential ($M_{\perp a}$, $K_a$) to ${\mathbb S}^2$ & (\ref{def:NaMaKa}) \\ \hline
$T_{ab}^\alpha$ & differential operators generating $\Psi_{\JM ab}^\alpha$ from $\Psi_{\JM}$&
(\ref{def:Tab}) \\ \hline
$W_{ab}^\alpha$ & differential operators generating $Y_{\JM ab}^\alpha$ from $Y_{\JM}$&
(\ref{def:Wab}) \\ \hline
$T^J_{\alpha l}$, $T^J_{\lambda \alpha}$, $T^J_{\lambda l}$ &
transformation matrices between different vector bases &
(\ref{eqn:vectortransformations})  \\ \hline
$U^J_{\alpha l}$, $U^J_{\lambda \alpha}$, $U^J_{\lambda l}$ &
transformation matrices between different tensor bases &
(\ref{eqn:tensortransformations})  \\ \hline
$\Delta_a$ &  lensing deflection field &
(\ref{eqn:deflection-definition}) \\ \hline
$\Pi_{ab}$ &  projection tensor &
(\ref{eqn:vectorsphericalharmonics}) \\ \hline
$\eta$ and $\eta_0$ & conformal time and its value today &
(\ref{eqn:deflection-definition}) \\ \hline
$\varphi(\hatn)$ and $\Omega(\hatn)$ & projected lensing potentials &
(\ref{eqn:deflectionbreakdown}) \\ \hline
$\varphi_{(JM)}$ and $\Omega_{(JM)}$ & lensing-potential
spherical-harmonic coefficients &
(\ref{eqn:potentialpowerspectra}) \\ \hline
$C_J^{\varphi\varphi}$ and $C_J^{\Omega\Omega}$ & lensing
angular power spectra & (\ref{eqn:potentialpowerspectra}) \\ \hline
$T^{\rm sca}(k)$ & density-perturbation transfer function &
(\ref{eqn:metricperturbation}) \\ \hline
$a(\eta)$ and $D_1(\eta)$ & scale factor and
density-perturbation growth factor &
(\ref{eqn:metricperturbation}) \\ \hline
$\Phi^{k,p}_{(JM)}$ & primordial amplitude for TAM wave of
density perturbation & (\ref{eqn:metricperturbation}) \\ \hline
$F_J^{\rm sca,GW}(k)$ &  lensing transfer function for density
perturbations and gravitational waves & (\ref{eqn:scalenstrans}) \\ \hline
$C_J^{EE}$ and $C_J^{BB}$ & lensing angular power spectra &
(\ref{eqn:Epowerspectrum}) \\ \hline
$T(k,\eta)$ &  time evolution for gravitational wave &
(\ref{eqn:GWpert}) \\ \hline
$f_J^X(kr)$ & radial eigenfunctions for gravitational-wave
transfer functions & (\ref{eqn:radialfunctions}) \\ \hline
 $\Theta(\bfx,\hatq;\eta)$ & radiation perturbation &
 (\ref{eqn:radiationpert}) \\ \hline
$\Theta^{k,JM}_{ll'}$ & expansion coefficients for radiation
perturbation &  (\ref{eqn:radiationpert}) \\ \hline
$\Xi_{ll'}^{k,JM}(\bfx,\hatq)$ & TAM eigenfunctions of $\bfx$
and $\hatq$ & (\ref{eqn:radiationpert}) \\ \hline
$\hatq$ & direction of photon momentum &
(\ref{eqn:radiationpert}) \\ \hline
$\mathcal{O}_m^l$ & $m$th component of irreducible-tensor
operator of rank $l$ & (\ref{eq:proof-objective}) \\ \hline
$\mathcal{O}_a^\alpha$ & vector operator for $\alpha=L,E,B$ &
(\ref{eqn:ibp}) \\ \hline
$\mathcal{D}^l_{mm'}$ &  Wigner rotation matrix &
(\ref{eqn:wignerrotation}) \\ \hline
$\mathcal{R}$ & an $O(3)$ rotation & (\ref{eq:proof-objective})
\\ \hline
\end{tabular}
\caption{A list of mathematical symbols used. The number in
the right-most column indicates the equation in or near which
the symbol is first defined or used.}
\end{center}
\end{table}

\section{Scalar fields}
\label{sec:scalarfields}

\subsection{Plane waves}
\label{sec:scalarplanewaves}

We begin with scalar fields to provide a simple introduction.
Our aim is to find solutions $\phi(\bfx)$ to the scalar
Helmholtz equation, $(\nabla^2 + k^2)\phi(\bfx)=0$.
The most general solution can be written in terms of plane waves
$\Psi^{\bfk}(\bfx) = e^{ i \bfk \cdot \bfx}$, eigenfunctions of
the momentum operator $-i\nabla$.  The set of solutions for all $\bfk$
constitute a complete orthonormal basis for scalar functions
$\phi(\bfx)$, normalized so that
\begin{equation}
     \int\, d^3 x \, \Psi^{\bfk}(\bfx) \left[\Psi^{\bfk'}(\bfx) \right]^*
     = (2\pi)^3 \delta_D(\bfk-\bfk'),
\label{def:Phibfk}
\end{equation}
where $\delta_D(\bfk-\bfk')$ is a Dirac delta function.
The most general scalar function can then be expanded,
\begin{equation}
     \phi(\bfx) = \int\, \frac{d^3k}{(2\pi)^3} \tilde \phi(\bfk)
     \Psi^{\bfk}(\bfx), \qquad \text{where} \quad
     \tilde\phi(\bfk) = \int\, d^3x \phi(\bfx) \left[
     \Psi^{\bfk}(\bfx) \right]^*.
\label{def:phixphik}
\end{equation}
The power spectrum $P(k)$ for a scalar field is then defined by
\begin{equation}
     \VEV{ \tilde\phi(\bfk) \tilde \phi^*(\bfk')} = (2\pi)^3
     \delta_D(\bfk-\bfk') P(k),
\label{def:pk_scalar}
\end{equation}
where the angle brackets denote an expectation value over all
realizations of the random field.  

\subsection{Total-angular-momentum waves}

Our aim here, though, is to find solutions that are
eigenfunctions of angular momentum.  This is easily done
with the plane-wave expansion,
\begin{equation}
     e^{i\bfk\cdot \bfx} = \sum_{lm} 4\pi i^l j_l(kr)
     Y_{(lm)}^*(\hatk) Y_{(lm)}(\hatn),
\label{eq:Rayleigh}
\end{equation}
where $j_l(x)$ is a spherical Bessel function, and $Y_{(lm)}(\hatn)$
are (scalar) spherical harmonics.\footnote{We choose the
spherical Bessel function of the first kind $j_l(kr)$ rather
than the second kind $n_l(kr)$, so that the TAM waves are regular at the
origin.  There may be cases, for example in application of this
formalism to emission or scattering of gravitational radiation,
in which the second function $n_l(kr)$ may need to be
introduced.}
We then find that if we
choose total-angular-momentum (TAM) basis functions,
\begin{equation}
     \Psi_{(lm)}^k(\bfx) \equiv j_l(kr)
     Y_{(lm)}(\hatn),
\label{def:Psilmk}
\end{equation}
where $r\equiv |\bfx|$ and $\hatn \equiv \bfx/r$, then an
arbitrary scalar function can be expanded as
\begin{equation}
     \phi(\bfx) = \sum_{lm} \int \frac{ k^2\, dk}{(2\pi)^3} \phi_{(lm)}(k)
     4 \pi i^l \Psi_{(lm)}^k(\bfx),
\end{equation}
with
\begin{equation}
     \phi_{(lm)}(k) = \int d^3 \bfx\,
     \left[4 \pi i^l \Psi_{(lm)}^k(\bfx)\right]^* \phi(\bfx)
     = \int\, d^2\hatk\, \tilde \phi(\bfk) Y_{(lm)}^*(\hatk).
\label{eqn:philmdef}
\end{equation}
Here we have used the relations,
\begin{equation}
     \int\, k^2 \, dk j_l(kr) j_l(kr') = \frac{\pi}{2 r^2}
     \delta_D(r-r'), \qquad
     \sum_{lm} Y_{(lm)}(\hatn) Y_{(lm)}^*(\hatn') =
     \delta_D(\hatn-\hatn').
\label{def:norm_jl_Ylm}
\end{equation}
The orthonormality relation for the basis functions is
\begin{equation}
     16\pi^2 \int\ d^3x\, \left[\Psi_{(lm)}^k(\bfx)\right]^*
     \Psi_{(l'm')}^{k'}(\bfx)
     =\delta_{ll'}\delta_{mm'} \frac{(2\pi)^3}{k^2}
     \delta_D(k-k'),
\end{equation}
where $\delta_{ij}$ is the Kronecker delta.
The basis functions also satisfy,
\begin{equation}
     \sum_{lm} \int \frac{k^2\, dk}{(2\pi)^3} \left [4 \pi i^l
     \Psi^k_{(lm)} (\bfx) \right]^* \left [4 \pi i^l
     \Psi^k_{(lm)} (\bfx') \right] = \delta_D(\bfx-\bfx'),
\end{equation}
which demonstrates that the $\Psi^k_{(lm)}(\bfx)$ constitute a
complete basis for scalar functions on $\mathbb{R}^3$.
The products of TAM-wave coefficients have expectation values,
\begin{equation}
     \VEV{ \phi_{(lm)}(k) \phi_{(l'm')}^*(k') } = \frac{(2\pi)^3}{k^2}
     \delta_D(k-k') \delta_{ll'} \delta_{mm'} P(k).
\end{equation}

\section{Vector fields}
\label{sec:vectorfields}

\subsection{Plane waves}
\label{sec:vectorplanewaves}

We now generalize to vector fields.  Again we begin by reviewing
plane-wave vector solutions to the Helmholtz equation.  
Three solutions to the vector Helmholtz equation, $(\nabla^2
+k^2)\Psi_a(\bfx)=0$ can be obtained, for each Fourier wavevector $\bfk$,
as
\begin{eqnarray}
     \Psi_a^{L,\bfk}(\bfx) &=& (1/k)\nabla_a \Psi^{\bfk}(\bfx)=
     i\hat k_a e^{i\bfk\cdot\bfx}, \nonumber\\
     \Psi_a^{1,\bfk}(\bfx) &=&  \frac{1}{|\bfk\times \hatz|}
     \left(\nabla \times \hatz \Psi^{\bfk}(\bfx) \right)_a = \frac{1}{|\bfk\times \hatz|}
     \epsilon_{abc} \nabla^b \hat z^c \Psi^{\bfk}(\bfx) =
     \frac{i \left[\bfk\times \hatz\right]_a}{|\bfk\times \hatz|}  e^{i \bfk \cdot \bfx}, \nonumber \\
     \Psi^{2,\bfk}_a(\bfx) &=& \frac{-i}{k}
     \epsilon_{ab}{}^c \,\nabla^b
     \Psi^{1,\bfk}_c(\bfx) = \frac{-i}{k |\bfk\times \hatz|} (\nabla_a \nabla_b -g_{ab} \nabla^2) \hat
     z^b \Psi^{\bfk}(\bfx) =\frac{i \left[ \bfk \times  (\bfk \times
     \hatz) \right]_a}{k |\bfk\times \hatz|} e^{i \bfk \cdot \bfx},
\label{eqn:vectorFourier}
\end{eqnarray}
where $\hatz$ is a unit vector in the $z$ direction, and
$\epsilon_{abc}$ is the totally antisymmetric tensor.
Here $\Psi^{L,\bfk}_a(\bfx)$ is a longitudinal vector field, and $\Psi^{1,\bfk}_a(\bfx)$ and 
$\Psi^{2,\bfk}_a(\bfx)$ are the two linear polarizations for the
transverse part of the vector field.  We could have written
Eq.~(\ref{eqn:vectorFourier}) more simply as
$\Psi_a^{X,\bfk}(\bfx) = i \hat\varepsilon^X_a(\hatk) e^{i\bfk
\cdot\bfx}$, with $\hat\varepsilon^L_a = k_a$ and
$\hat\varepsilon^{1,2}_a$ two other unit vectors orthogonal to
$k_a$ and to each other.  We have written in
Eq.~(\ref{eqn:vectorFourier}) one choice for these
polarization vectors explicitly in terms of a fixed unit vector
$\hat z$ to motivate a choice of polarization vectors for the
TAM waves later.

These mode functions are normalized so that they constitute a
complete orthonormal set,
\begin{equation}
     \int \, d^3x\, \Psi^{\alpha,\bfk\,a}(\bfx)
     \left[\Psi^{\beta,\bfk}_{\,\,a}(\bfx)\right]^* = (2\pi)^3
     \delta_D(\bfk-\bfk') \delta_{\alpha\beta},
\end{equation}
where $\alpha,\beta=\{L,1,2\}$.  The three mode functions are,
furthermore, orthogonal at each point.  An arbitrary vector
field $V_a(\bfx)$ can then be expanded as
\begin{equation}
     V_a(\bfx) = \int \frac{d^3k}{(2\pi)^3} \left[ \tilde
     V^L(\bfk) \Psi^{L,\bfk}_a(\bfx) + \tilde
     V^1(\bfk) \Psi^{1,\bfk}_a(\bfx) + \tilde
     V^2(\bfk) \Psi^{2,\bfk}_a(\bfx) \right],
\label{eqn:vectorexpansion}
\end{equation}
in terms of Fourier expansion coefficients,
\begin{eqnarray}
     \tilde V^L(\bfk) = \int \, d^3x\, V^a(\bfx)
     \left[\Psi^{L,\bfk}_a(\bfx)\right]^* = -\int\,
     d^3x\,\left[\Psi^{\bfk}(\bfx) \right]^* \frac{1}{k}
     \nabla^a  V_a(\bfx),
     \nn \\
     \tilde V^1(\bfk) = \int \, d^3x\, V^a(\bfx)
     \left[\Psi^{1,\bfk}_a(\bfx)\right]^* = \int\, d^3x\,
     \left[\Psi^{\bfk}(\bfx) \right]^*
     \frac{1}{| \bfk \times \hatz |} \epsilon_{abc} \hat z^a
     \nabla^b V^c(\bfx),
     \nn \\
     \tilde V^2(\bfk) = \int \, d^3x\, V^a(\bfx)
     \left[\Psi^{2,\bfk}_a(\bfx)\right]^* = \int\, d^3x\,
     \left[\Psi^{\bfk}(\bfx)\right]^*
     \frac{-i}{k| \bfk \times \hatz |}  \hat{z}^a 
     (\nabla_a \nabla_b - g_{ab} \nabla^2) V^b(\bfx).
\end{eqnarray}
We obtain the last equality in each line by integrating by
parts.

The statistics of the vector field are given in terms of power
spectra $P_L(k)$ and $P_T(k)$ for the longitudinal and
transverse components, respectively, that satisfy
\begin{eqnarray}
     \VEV{ \tilde V^L(\bfk) \tilde V^{L\,*}(\bfk')} &=& (2\pi)^3
     \delta_D(\bfk - \bfk') P_L(k), \nonumber \\
     \VEV{ \tilde V^1(\bfk) \tilde V^{1\,*}(\bfk')} &=& (2\pi)^3
     \delta_D(\bfk - \bfk') P_T(k), \nonumber \\
     \VEV{ \tilde V^2(\bfk) \tilde V^{2\,*}(\bfk')} &=& (2\pi)^3
     \delta_D(\bfk - \bfk') P_T(k).
\label{def:pk_vector}
\end{eqnarray}
The decomposition of the transverse component into the two modes
$V^\alpha$ ($\alpha=1,2$) is not rotationally invariant---the decomposition
would be different if we had chosen a different direction for
$\hatz$---so the power spectra for the two must be the same.
However, we can alternatively decompose the transverse-vector
modes into plane waves of right and left circular polarization,
or positive and negative helicity,
\begin{equation}
     \Psi^{\pm,\bfk}_a(\bfx) 
		= \frac{1}{\sqrt{2}}\left( \Psi^{1,\bfk}_a(\bfx) \pm i 
     \Psi^{2,\bfk}_a(\bfx) \right).
\label{def:vectorPolarization}
\end{equation}
Since $\Psi^{2,\bfk}_a(\bfx) = -(i/k) \epsilon_{abc} \nabla^b
\Psi^{1,\bfk,c}(\bfx)$, these modes are invariant under rotations
about the $\hatk$ direction and thus in some sense
more ``physical'' than the 1 and 2 linear polarizations.  It is
possible (although it would require parity breaking)
that $P_+(k)$ and $P_-(k)$ could differ.  In the absence of
parity breaking $P_+(k)=P_-(k)=P_T(k)$.

\subsection{TAM Waves}
\label{sec:vectortamwaves}

The aim now is to find vector-valued functions $V_a(\bfx)$ that
satisfy the vector Helmholtz equation,
$\left(\mathbf{\nabla}^{2}+k^{2}\right)V_a(\bfx)=0$,
for definite wavenumber magnitude $k$, and that transform under
spatial rotation as representations of
order $J$. In other words, we seek eigenfunctions of total
angular momentum $\mathbf{J}=\mathbf{L}+\mathbf{S}$, where
$L_a=-i \epsilon_{abc} x^b \nabla^c$ is the orbital
angular momentum and $\mathbf{S}$ is the $S=1$ spin associated
with the vector space spanned by a set of basis vectors at each
spatial point.  This differs from the case of scalar fields
where, with spin $S=0$, total-angular-momentum eigenstates
coincide with orbital-angular-momentum states.

Our strategy will be to first construct vector-valued
eigenfunctions of total angular momentum that are also
eigenfunctions of orbital angular momentum $l$.  We will then
construct linear combinations of states of definite total
angular momentum $JM$ that are curl-free (the longitudinal
component) and divergence-free (the two transverse components).
We will then decompose the TAM waves also into helicity eigenstates.

\subsubsection{The orbital-angular-momentum basis}
\label{sec:OAMvector}

From the usual set of Cartesian basis unit vectors $e^\mu_a =
\delta^\mu_a$, for $\mu=x,y,z$, one can construct a spherical
basis $e^{\bar{m}}_a$ for $\bar{m}=+1,0,-1$,
through\footnote{Barred indices like $\bar{m}$ are
reserved for the order-1 spherical basis.}
\be
     e^0_a =e^z_a, \quad e^{\pm}_a =
     \mp\left(e^x_a\pm  i e^y_a\right)/\sqrt{2}.
\label{def:spherical_basis_vector}
\ee
These constitute a complex but global basis, so these unit
vectors commute with differential operators. Under spatial
rotations, they transform
as an $l=1$ representation. We know for the spatial part of the
eigenfunction that the conventional scalar-valued
spherical harmonics $Y_{(lm)}(\hatn)$ form a representation of
order $l$ of the spatial rotation group. Vector-valued
eigenfunctions of total angular momentum are therefore
constructed via the usual scheme for adding two angular
momenta \cite{Varshalovich},
\begin{equation}
     \Psi_{(JM)a}^{l,k}(\bfx)= j_l(kr)
     Y_{(JM)a}^l(\hatn)  \equiv
     \sum_{m\bar{m}}\langle1\bar{m}lm|JM\rangle
     j_{l}(kr) Y_{(lm)}(\hatn) e^{\bar{m}}_a,
\label{eqn:CBexpansion}
\end{equation}
where $\VEV{l_1 m_1 l_2 m_2|JM}$ are Clebsch-Gordan
coefficients.  Here, $J=0,1,2,\ldots$, $M=-J,-J+1,\ldots,J-1,J$,
and $l=J-1,J,J+1$.  The TAM waves $\Psi_{(JM)a}^{l,k}(\bfx)$ are
also eigenfunctions of orbital angular momentum squared
$\mathbf{L}^2 = L^a L_a$ with eigenvalue $l(l+1)$. The
angular parts $Y_{(JM)a}^l(\hatn)$ (three-dimensional vector
spherical harmonics of given total and orbital angular momentum)
are normalized to
\be
     \int
     d^2\hatn \left[Y_{(J'M')}^{l'\, a} (\hatn) \right]^*
     Y_{(JM)a}^l(\hatn) = \delta_{ll'}\delta_{JJ'}\delta_{MM'}.
\ee
There are three eigenfunctions for given total angular momentum
$JM$ distinguished by their orbital angular momentum $l$.  The
TAM waves are normalized so that
\begin{equation}
     \int \, d^3 x \, \left[ 4 \pi i^J \Psi^{l,k}_{(JM)a}(\bfx)
     \right]^* 4 \pi i^{J'} \Psi^{l',k'\, a}_{(J'M')}(\bfx) =
     \delta_{ll'}\delta_{JJ'}\delta_{MM}' \frac{(2\pi)^3}{k^2}
     \delta_D(k-k').
\label{eqn:vecsphernorm}
\end{equation}

We also have that
\ba
\sum_{JMl}
\int \frac{k^2dk}{(2\pi)^3}
\left[
4\pi i^l \Psi_{\JM}^{l,k\,\,a}(\bfx')
\right]^*
\left[
4\pi i^l \Psi_{\JM a}^{l,k}(\bfx)
\right] = \delta_D(\bm{x}-\bm{x}'),
\label{eqn:vectorcompleteness}
\ea
which demonstrates that the $\Psi^{l,k}_{(JM)a}(\bfx)$
constitute a complete basis for vector functions on
$\mathbb{R}^3$.  To show this, we use the definition in
Eq.~(\ref{eqn:CBexpansion}) to rewrite the left-hand side as
\be
\sum_{JMl}
\int \frac{k^2dk}{(2\pi)^3}
(4\pi)^2 j_l(kr)j_l(kr')  
\left[
Y_{\JM}^{l\,\,a}(\hatn')
\right]^*
Y_{\JM a}^l(\hatn)
=
\left[
\frac{2}{\pi}
\int k^2dk
j_l(kr)j_l(kr')  
\right]
\sum_{JMl}
\left[
Y_{\JM}^{l\,\,a}(\hatn')
\right]^*
Y_{\JM a}^l(\hatn).
\ee
The $k$ integral is 
\be
\frac{2}{\pi}
\int k^2dk
j_l(kr)j_l(kr')  =  \frac1{r^2} \delta_D(r-r'),
\ee
and the sum then becomes
\ba
\sum_{JMl}
\left[
Y_{\JM}^{l\,\,a}(\hatn')
\right]^*
Y_{\JM a}^l(\hatn)
&=&
\sum_{JMl}
\sum_{m\bar{m}}\sum_{m'\bar{m}'}
\left<1\mbar lm|JM\right>
\left<1\mbar' lm'|JM\right>
\left[Y_{\lm}(\hatn') {e}^{\bar{m}\,a}\right]^* 
Y_{(lm')}(\hatn){e}_a^{\bar{m}'}
\nn\\
&=&
\sum_{JMl}
\sum_{m\bar{m}}\sum_{m'\bar{m}'}
\left<1\mbar lm|JM\right>
\left<1\mbar' lm'|JM\right>
\left[Y_{\lm}(\hatn') \right]^* 
Y_{(lm')}(\hatn)\delta_{\bar{m}\bar{m}'}
\nn\\
&=&
\sum_{lm\bar{m}}\sum_{m'\bar{m}'}
\left[Y_{\lm}(\hatn') \right]^* 
Y_{(lm')}(\hatn)\delta_{\bar{m}\bar{m}'}
\delta_{mm'}
\nn\\
&=&
\sum_{lm}
\left[Y_{\lm}(\hatn') \right]^* 
Y_{(lm')}(\hatn)
=
\delta_D(\hatn-\hatn'),
\ea
from which Eq.~(\ref{eqn:vectorcompleteness}) follows.  Note
that this also demonstrates that the $Y^l_{(JM)a}(\hatn)$
constitute a complete basis for three-dimensional vectors on the
two-sphere.

\subsubsection{The longitudinal/transverse basis}

The next step will be to construct linear combinations of these
OAM waves that are longitudinal and transverse.
To do so, we must calculate the divergence and curl of
$\Psi_{(JM)a}^{l,k}(\bfx)$.  The result for the divergence,
detailed in Appendix \ref{appendix:vectordivergence}, is
\be
\label{eq:div}
     r\nabla^a \Psi^l_{(JM)a}(\bfx)=\begin{cases}
     -\sqrt{\frac{J}{2J+1}}\left(kr\right) j_{J}(kr) Y_{\JM}(\hatn), & \qquad
     l=J-1,\\ 0, & \qquad l=J,\\
     -\sqrt{\frac{J+1}{2J+1}}\left(kr\right) j_{J}(kr) Y_{\JM}(\hatn), &
     \qquad l=J+1.\end{cases}
\ee
Since the parity of a given OAM state is
$(-1)^l$, and the basis vectors in Eq.~(\ref{eqn:CBexpansion})
of odd parity, we choose transverse-vector fields of parity
$(-1)^{J}$ and $(-1)^{J+1}$ to be, respectively,\footnote{The
parity of $(-1)^J$ for the vector $B$ mode and $(-1)^{J+1}$ for
the vector $E$ mode differ by $-1$ from the parities of the
$E/B$ tensor spherical harmonics.  However, the expansion
coefficients for the vector $E$ and $B$ modes have, as we will
see below, parities $(-1)^{J+1}$ and $(-1)^J$, as do the
tensor-spherical-harmonic expansion coefficients.  The reason
traces back to the transformation property of the vector field
under a parity inversion.}
\begin{equation}
     \Psi^{B}_{(JM)a}(\bfx) = i \Psi^J_{(JM)a}(\bfx), \qquad \text{and} \qquad
     \Psi^E_{(JM)a}(\bfx) = i\left[\sqrt{\frac{J+1}{2J+1}} \Psi^{J-1}_{(JM)a}(\bfx)
     -\sqrt{\frac{J}{2J+1}} \Psi^{J+1}_{(JM)a}(\bfx) \right].
\end{equation}
The basis functions for the longitudinal field may then be taken
to be
\begin{equation}
     \Psi^L_{(JM)a}(\bfx) = i\left[ \sqrt{\frac{J}{2J+1}} \Psi^{J-1}_{(JM)a}(\bfx)
     + \sqrt{\frac{J+1}{2J+1}} \Psi^{J+1}_{(JM)a}(\bfx) \right].
\end{equation}
The prefactors have been chosen so that the three sets of
eigenfunctions are normalized as in
Eq.~(\ref{eqn:vecsphernorm}).  
We thus have a complete orthonormal set of basis functions, of fixed
total angular momentum, for the transverse and longitudinal
components of a vector field.

\subsubsection{The longitudinal/transverse basis in terms of
derivative operators}
\label{sec:vectoroperators}

There is, however, an alternative and useful route to these
longitudinal and transverse basis functions.  In
Appendix~\ref{appendix:irreducible} it is proved that if an
operator $\mathcal{O}$ is an irreducible tensor under rotations, then $J^2
\mathcal{O} Y_{(JM)} = J(J+1) \mathcal{O} Y_{(JM)}$ and $J_z
\mathcal{O} Y_{(JM)} = M \mathcal{O} Y_{(JM)}$.  We can
therefore construct vector TAM waves by applying appropriately
defined vector operators to scalar TAM waves.

Consider three vector operators 
\be
D_a\equiv\frac{i}{k}\nabla_a,\quad K_a\equiv -iL_a,\quad
M_a\equiv\epsilon_{abc}D^b K^c.
\label{eqn:operatordefs}
\ee
These are irreducible-vector operators, and they all commute
with $\nabla^2$. They therefore yield, when acting on scalar TAM
waves, TAM vector waves of total angular momentum $JM$ that are
also solutions of the vector Helmholtz equation.  These three
sets of vector fields must be linear combinations of
$\Psi^{l,k}_{(JM)}(\bfx)$, for $l=J-1,J,J+1$ or
$\Psi^{\alpha,k}_{(JM)}(\bfx)$ for $\alpha=L,E,B$. 
Since the three operators satisfy
\be
D^a K_a=K^a D_a=0,\qquad D^a M_a=0,\qquad M^a D_a=2,\qquad K^a
M_a = M^a K_a =0, 
\ee
it follows that $D_a$ generates the longitudinal vector field
$D_a\Psi^{k}_{(JM)}(\bfx)\propto\Psi^{L,k}_{(JM)a}(\bfx)$, while $K_a$ and $M_a$
generate divergence-free vector fields.  Since $K_a$ is
axial-vector-like and $M_a$ vector-like, parity considerations
tell us that $K_a$, generates the $B$ mode,
$K_a\Psi^{k}_{(JM)}(\bfx)\propto\Psi^{B,k}_{(JM)a}(\bfx)$, while $M_a$
generates the $E$ mode,
$M_a\Psi^{k}_{(JM)}(\bfx)\propto\Psi^{E,k}_{(JM)a}(\bfx)$.

The operators $D_a$, $K_a$, and $M_a$ are also operators in the
Hilbert space of vector-valued fields, so we can calculate their
hermitian conjugates to be
\be
     \left(D_{a}\right)^{\dagger} = D_{a}, \qquad
     \left(K_{a}\right)^{\dagger} = -K_{a}, \qquad
     \left(M_{a}\right)^{\dagger}=-M_{a}+2D_{a}.
\label{eqn:DKM_conjugate}
\ee
Thus, when acting on $\Psi^{k}_{(JM)}$, the three operators have norms
\be
     (D^a)^{\dagger} D_a=1,\qquad (K^a)^{\dagger} K_a =L^a
     L_a=J(J+1),\qquad (M^a)^{\dagger} M_a =L^a L_a=J(J+1).
\ee
These results enable us to normalize the vector TAM waves and to
reproduce the longitudinal/transverse basis. This operator
approach has the advantage that many calculations involving
vector or higher-spin TAM waves can be reduced to the
algebra of operators that act on scalar spherical waves. The
following properties of the three operators will be useful
in calculations:
\begin{align}
    &\left[D_a,D_b\right] = 0, \qquad \left[K_{a},D_{b}\right] =
    \epsilon_{abc}D^{c}, \qquad \left[M_{a},D_{b}\right]
    =g_{ab} -D_{a}D_{b},\nn\\
    &\left[K_{a},K_{b}\right] = \epsilon_{abc}K^{c}, \qquad
    \left[K_{a},M_{b}\right] = \epsilon_{abc}M^{c}, \qquad
    \left[M_{a},M_{b}\right]=-\epsilon_{abc}K^{c}.
\end{align}
We can gain insight into the operators $D_a$, $K_a$, and $M_a$
from the far-field limit $kr\rightarrow\infty$, where $D_a$ is
approximated by an ordinary vector in the radial direction, and
$K_a$ and $M_a$ asymptote to two orthogonal vectors in the plane
perpendicular to the radial direction, when they act on a scalar
TAM wave. 
The factor of $i$ in the definition of $D_a$ is chosen so that
$D_a=-\hat{k}_a$ in this limit.  The sign convention for the $E$/$B$
vector TAM waves is chosen so that if we rotate the $E$ mode by
$+90^{\circ}$ about the direction of wave propagation we obtain
a $B$ mode.

To summarize, the decomposition into longitudinal and transverse
modes is
\begin{align}
     \mathrm{B\, mode:} &
     \quad\Psi^B_{(JM)a}(\bfx) = \frac{K_a}{\sqrt{J(J+1)}} \Psi_{(JM)}(\bfx)
     = \frac{-i}{k} \epsilon_{abc} \nabla^b
     \Psi^{E\,\,\,c}_{(JM)}(\bfx) =i \Psi^J_{(JM)a}(\bfx),
     \nonumber \\
     \mathrm{E\, mode:} &
     \quad\Psi^E_{(JM)a}(\bfx) = \frac{M_a}{\sqrt{J(J+1)}} \Psi_{(JM)}(\bfx)
     = \frac{i}{k} \epsilon_{abc} \nabla^b
     \Psi^{B\,\,\,c}_{(JM)}(\bfx) \nn \\
     & \quad \;\;  \qquad \qquad =
     i\left[ \left(\frac{J+1}{2J+1}\right)^{1/2}
     \Psi^{J-1}_{(JM)a}(\bfx) - \left(\frac{J}{2J+1}\right)^{1/2}
     \Psi^{J+1}_{(JM)a}(\bfx) \right], \nn\\
     \mathrm{longitudinal\, mode:}& \quad\Psi^L_{(JM)a}(\bfx) =
     D_a  \Psi_{(JM)}(\bfx) =
     i \left[ \left(\frac{J}{2J+1}\right)^{1/2} \Psi^{J-1}_{(JM)a}(\bfx) +
       \left(\frac{J+1}{2J+1}\right)^{1/2} \Psi^{J+1}_{(JM)a}(\bfx)
       \right].
\label{eqn:vectorTAMs}
\end{align}

\subsubsection{The helicity basis}

We can define another basis, denoted by the helicity
$\lambda=0,\pm1$, by
\begin{equation}
     \Psi^{\pm1}_{(JM)a}(\bfx) = \frac{1}{\sqrt{2}}
     \left[\Psi^E_{(JM)a}(\bfx) \pm
     i\Psi^B_{(JM)a}(\bfx) \right], \qquad
     \Psi^{0}_{(JM)a}(\bfx) = \Psi^L_{(JM)a}(\bfx).
\label{def:Psialambda}
\end{equation}
These are eigenstates of the helicity operator $H={\bf S}\cdot
{\bf{\hat p}}$, where $(S_b)_{ac} = i \epsilon_{abc}$ is the
spin operator and ${\hat{p}_a} = -i\nabla_a/k$ the normalized
momentum operator, with eigenvalues $\lambda$.

We may summarize the transformation between the three
bases---labeled by $l=J,J-1,J+1$ for the
orbital-angular-momentum basis, $\alpha=B,E,L$ for the
longitudinal/transverse basis, and $\lambda = 1,0,-1$ for the
helicity basis---by the transformation matrices,
\begin{equation}
     T^J_{\alpha l} = i \left( \begin{array}{ccc}  1 & 0 & 0 \\ 0 &
     \sqrt{\frac{J+1}{2J+1}} &  - \sqrt{\frac{J}{2J+1}} \\
     0 & \sqrt{\frac{J}{2J+1}} &  
     \sqrt{\frac{J+1}{2J+1}} \end{array}\right), \qquad
     T^J_{\lambda \alpha} = \left( \begin{array}{ccc}
     \frac{i}{\sqrt{2}} & \frac{1}{\sqrt{2}} & 0 \\
     0 & 0 & 1 \\
     \frac{-i}{\sqrt{2}} & \frac{1}{\sqrt{2}} & 0 \end{array} \right), \qquad
     T^J_{\lambda l} = i \left(\begin{array}{ccc}
     \frac{i}{\sqrt{2}} & \sqrt{\frac{J+1}{2\left(2J+1\right)}} & -\sqrt{\frac{J}{2\left(2J+1\right)}}\\
     0 & \sqrt{\frac{J}{2J+1}} & \sqrt{\frac{J+1}{2J+1}}\\
     \frac{-i}{\sqrt{2}} & \sqrt{\frac{J+1}{2\left(2J+1\right)}} & -\sqrt{\frac{J}{2\left(2J+1\right)}}
     \end{array}\right).
\label{eqn:vectortransformations}
\end{equation}

\subsubsection{Projection onto Vector Spherical Harmonics}
\label{sec:vectorprojection}

Here we have constructed three different bases for
three-dimensional vectors on ${\mathbb R}^3$.  We now show how
the angular components project onto the more familiar vector
spherical harmonics $Y^E_{(JM)a}(\hatn)$ and
$Y^B_{(JM)a}(\hatn)$, for two-dimensional vectors that live on
the two-sphere.  These vector spherical harmonics are given by
\begin{equation}
    Y^E_{(JM)a}(\hatn) = \frac{-r}{\sqrt{J(J+1)}} \nabla_{\perp\,a}
    Y_{(JM)}(\hatn), \qquad     Y^B_{(JM)a}(\hatn) =
    \frac{-r}{\sqrt{J(J+1)}} \epsilon_{abc} \hat{n}^b \nabla^c
    Y_{(JM)}(\hatn),
\label{eqn:vectorsphericalharmonics}
\end{equation}
both of which have $\hat n^a Y^E_{(JM)a}(\hatn) = 0 = \hat n^a
Y^B_{(JM)a}(\hatn)$.  Here $\nabla_{\perp\,a}=\Pi_a{}^b\nabla_b$
is the gradient operator in the $\hat\theta$-$\hat\phi$ space,
and $\Pi_{ab} = g_{ab}-\hat n_a \hat n_b$ projects onto that space.
In addition, we can define a third vector
spherical harmonic $Y^L_{(JM)a}(\hatn) = -\hat n_a Y_{(JM)}(\hatn)$
to account for the component of a three-dimensional vector in
the normal direction.

This set of three vector spherical harmonics provides a complete
set of orthonormal basis functions for three-dimensional
vectors that live on the two-sphere.  We can obtain these vector
spherical harmonics using an operator approach that parallels
that we developed for TAM waves.  Define three dimensionless
irreducible-vector operators,
\be
N_a=-\hat{n}_a,
\qquad K_a=-iL_a,
\qquad  M_{\perp
a}=\epsilon_{abc}N^bK^c.
\label{def:NaMaKa}
\ee 
These are analogues of the three operators
$D_a$, $K_a$, and $M_a$ we defined to derive vector TAM waves,
but they act on the Hilbert space of all functions of
$\hatn$; i.e. they do not act on the radial coordinate $r$. 
This new set of operators satisfies precisely the same algebra as 
the set $\{D_a,K_a,M_a\}$. They are orthogonal to each
other,
\be
     N_{a}K^{a}=K_{a}N^{a} =0, \qquad N_{a}M_{\perp}^{a}=0,
     \quad M_{\perp a}N^{a}=2,\qquad K_{a}M_{\perp}^{a}=M_{\perp
     a}K^{a}=0,
\ee
and they are normalized to
\begin{equation}
     \left(N_{a}\right)^{\dagger}N^{a}=N_{a}N^{a}=1,\qquad
     \left(K_{a}\right)^{\dagger}K^{a}=-K_{a}K^{a}=\mathbf{L}^{2},
     \qquad \left(M_{\perp a}\right)^{\dagger}M_{\perp}^{a}=-M_{\perp
     a}M_{\perp}^{a}=\mathbf{L}^{2}.
\end{equation}
As operators in the Hilbert space, their hermitian conjugates are
\be
     \left(N_{a}\right)^{\dagger}=N_{a},
     \qquad\left(K_{a}\right)^{\dagger}= -K_{a},\qquad\left(M_{\perp
     a}\right)^{\dagger}=-M_{\perp a}+2N_{a}. 
\ee
Furthermore, they satisfy the algebraic relations,
\begin{align}
     &\left[N_{a},N_{b}\right]=0,\qquad\left[M_{\perp
     a},N_{b}\right] = \left(g_{ab}-N_{a}N_{b}\right),
     \qquad\left[M_{\perp a},M_{\perp
     b}\right]=-\epsilon_{abc}K^{c},\nn\\ 
     &\left[K_{a},K_{b}\right] =
     \epsilon_{abc}K^{c},\qquad\left[K_{a},N_{b}\right] =
     \epsilon_{abc} N^{c}, \qquad\left[K_{a},M_{\perp b}\right]
     =\epsilon_{abc}M_{\perp}^{c}.
\end{align}

The two operators $K_a$ and $M_{\perp a}$ generate the two
transverse-vector spherical harmonics
$Y_{(JM)a}^{E}(\hatn)$ and $Y_{(JM)a}^{B}(\hatn)$, in terms of
$E/B$ modes, while $N_a$ generates the longitudinal vector
spherical harmonic $Y_{(JM)a}^{L}(\hatn)$ in the normal
direction.  In summary,
\begin{align}
     &
     Y_{(JM)a}^{B}(\hatn)=\frac{1}{\sqrt{J\left(J+1\right)}}K_{a}Y_{(JM)}(\hatn)
     = iY_{(JM)a}^{J}(\hatn),\nn\\ 
     & Y_{(JM)a}^{E}(\hatn)=\frac{1}{\sqrt{J\left(J+1\right)}} M_{\perp
     a}Y_{(JM)}(\hatn) = -\sqrt{\frac{J+1}{2J+1}}Y_{(JM)a}^{J-1}(\hatn) -
     \sqrt{\frac{J}{2J+1}}Y_{(JM)a}^{J+1}(\hatn),\nn\\
     & Y_{(JM)a}^{L}(\hatn)=N_{a}Y_{(JM)}(\hatn) =
     -\sqrt{\frac{J}{2J+1}}Y_{(JM)a}^{J-1}(\hatn) +
     \sqrt{\frac{J+1}{2J+1}}Y_{(JM)a}^{J+1}(\hatn),
\label{eqn:vectorYlms}
\end{align}
normalized so that
\begin{equation}
     \int \, d^2\hatn \, \left[Y^{\alpha,a}_{(JM)} \left( \hatn \right)\right]^*
     Y^{\beta}_{(J'M')a} \left( \hatn \right) =\delta_{JJ'}
     \delta_{MM'} \delta_{\alpha\beta},
\end{equation}
for $\alpha,\beta=\{E,B,L\}$.  Note that although the relations
between the OAM and $E/B/L$ vector-spherical-harmonic bases in
Eq.~(\ref{eqn:vectorYlms}) resemble those between the OAM and
$E/B/L$ bases for TAM waves in Eqs.~(\ref{eqn:vectorTAMs}) and
(\ref{eqn:vectortransformations}), there are subtle, and
important, sign differences.
The minus sign in the definition of $N_a$ is chosen to match
$D_a=-\hat{k}_a$. The sign convention for E/B vector spherical
harmonics is chosen so that a rotation of the E mode by
$+90^{\circ}$ about the outward normal direction (i.e. the
direction of $\hatn$) yields the B mode.

Finally, we can write the $E/B/L$ TAM waves in terms of the
$E/B/L$ spherical harmonics:
\begin{eqnarray}
     \Psi^{k,B}_{(JM)a}(\bfx) &=& j_J(kr) Y^B_{(JM)a}(\hatn), \nn \\
     \Psi^{k,E}_{(JM)a}(\bfx) &=& -i\left[ j_J'(kr)
     +\frac{j_J(kr)}{kr} \right] Y^E_{(JM)a}(\hatn) -
     i\sqrt{J(J+1)} \frac{j_J(kr)}{kr} Y^L_{(JM)a}(\hatn), \nn
     \\
     \Psi^{k,L}_{(JM)a}(\bfx) &=& -i\sqrt{J(J+1)}
     \frac{j_J(kr)}{kr} Y^E_{(JM)a}(\hatn) - ij_J'(kr)
     Y^L_{(JM)a}(\hatn).
\label{eqn:vectorprojections}
\end{eqnarray}
Although the mode functions are orthonormal, we now see that
they are {\it not} orthogonal at each point.  Although the $L$
and $B$ modes are everywhere perpendicular and the $E$ and $B$
modes everywhere perpendicular, the $L$ and $E$ vector TAM waves are 
not always perpendicular.   The $B$ mode has components only in
the $\theta$-$\phi$ plane; i.e., $n^a \Psi^B_{(JM)a}(\bfx)=0$.  The
$E$ and $L$ modes most generally have components in the
tangential plane and along the normal $n^a$. In the far-field
limit $kr\rightarrow\infty$, however, the three modes are
asymptotically perpendicular to each other.

\subsubsection{The plane wave expansion for vector fields}
\label{sec:vectortransforms}

We now determine the transformation between the vector
plane-wave basis and the vector TAM-wave bases.  We start
with the OAM basis.  Since the $\Psi^{l,k}_{(JM)a}(\bfx)$ constitute a
complete basis, we may write,
\begin{equation}
     \hat \varepsilon_a(\bfk) e^{i \bfk \cdot \bfx} = \sum_{lJM} 4\pi i^l
     A_{(JM)}^l(\hatk) \Psi^{l,k}_{(JM)a}(\bfx)= \sum_{lJM} 4\pi i^l
     A_{(JM)}^l(\hatk) j_l(kr) Y^l_{(JM)a}(\hatn).
\label{eqn:vectorplanewaves}
\end{equation}
Here $\hat \varepsilon_a$ is a (unit) polarization vector for the
wave.  The coefficients $A_{(JM)}^l(\bfk)$ may be obtained by
writing $e^{i \bfk\cdot \bfx}$ in the usual scalar plane-wave
expansion, 
\begin{equation}
     \hat \varepsilon_a(\bfk) e^{i \bfk \cdot \bfx} = \sum_{lm} 4\pi
     i^l j_l(kr) Y_{(lm)}^*(\hatk) Y_{(lm)}(\hatn)
     \hat \varepsilon_a(\bfk).
\end{equation}
We then use orthonormality of the $Y_{(JM)a}^l(\hatn)$ to infer
that
\begin{equation}
     A_{(JM)}^l(\hatk) = \hat\varepsilon^a (\bfk) Y^{l\,\,*}_{(JM)a}(\hatk).
\end{equation}

We can similarly expand in terms of $L,E,B$ modes, or
helicity modes, as
\begin{equation}
     \hat\varepsilon_a(\bfk) e^{i \bfk \cdot \bfx} =
     \sum_{\alpha=L,E,B} \sum_{JM} 4\pi i^{J}
     A_{(JM)}^\alpha(\hatk) \Psi^{k,\alpha}_{(JM)a}(\bfx), \qquad
     \hat\varepsilon_a(\bfk) e^{i \bfk \cdot \bfx} =
     \sum_{\lambda=-1,0,1}\sum_{JM} 4\pi i^J
     A_{(JM)}^\lambda(\hatk) \Psi^{k,\lambda}_{(JM)a}(\bfx),
\end{equation}
in terms of expansion coefficients
\begin{equation}
     A_{(JM)}^\alpha(\hatk) = \hat\varepsilon^a (\bfk)
     Y^{\alpha\,*}_{(JM)a}(\hatk), \qquad
     A_{(JM)}^\lambda(\hatk) = \hat\varepsilon^a (\bfk)
     Y^{\lambda\,*}_{(JM)a}(\hatk).
\label{def:vector_helicity_Ylm}
\end{equation}
Here, the spin-1 vector spherical harmonics are
$Y^{\lambda=\pm1}_{(JM)a} = 2^{-1/2} \left[ Y^E_{(JM)a}
\pm i Y^B_{(JM)a} \right]$. These are
related to the usual spin-1 spherical harmonics
${}_{\lambda} Y_{(JM)}(\hatn)$ \cite{NewmanPenrose,Goldberg} by
\be
     \hat{\varepsilon}_{\lambda'}^a(\bfk)
     Y_{(JM)a}^{\lambda}(\mathbf{\hat{k}}) =
     {}_{-\lambda} Y_{(JM)}(\mathbf{\hat{k}})
     \delta_{\lambda\lambda^{\prime}},\qquad\lambda,\lambda^{\prime}=0,\pm1.
\ee
Here, $\hat\varepsilon^a_{\lambda}(\bfk)$, for $\lambda=0,\pm1$,
are the polarization vectors for a vector plane wave with
wavevector $\bfk$ and helicity $\lambda$. This equation defines
our phase convention for $\hat{\varepsilon}_{\lambda}$.
In terms of basis vectors in spherical coordinates, these are
defined as 
\be
\hat{\varepsilon}_0^a = \hat{n}^a, \qquad 
\hat{\varepsilon}_{\pm1}^a = \mp\frac{1}{\sqrt{2}}(\hat\theta^a
\mp i \hat\phi^a).
\label{eq:helicity_polarization_vector}
\ee

\subsubsection{Expansion of vector fields and power spectra}
\label{sec:vectorexpansion}

An arbitrary vector field $V_a(\bfx)$ can be
expanded in the OAM basis by
\begin{equation}
      V_a({\bf x}) = \sum_{JM} \sum_{l=J-1,J,J+1} \int
     \frac{k^2 dk}{(2\pi)^3} V_{(JM)}^l(k) 4\pi i^l
     \Psi^{l,k}_{(JM)a}(\bfx),
\label{eqn:vectorcoeffs}
\end{equation}
in terms of expansion coefficients,
\begin{equation}
     V_{(JM)}^l(k) = \int d^3 \bfx\, V^a(\bfx) \left[4
     \pi i^l \Psi_{(JM)a}^{l,k}(\bfx) \right]^*.
\end{equation}
As we show in Ref.~\cite{inprogress}, these can also be written
as vector-spherical-harmonic transforms,
\be
     V_{(JM)}^l(k) = \int d^2\hatk\, \tilde
     V^a(\bfk) Y^{l\,\,*}_{(JM)a}(\hatk),
\ee
of the vector Fourier coefficients.  Analogous relations hold
for the $L/E/B$ and helicity bases as well.

For the $L/E/B$ basis, we may use the operator
approach discussed above to rewrite the expansion coefficients
in terms of scalar TAM waves by integrating by parts:
\ba
V_{(JM)}^\alpha(k) 
&=& 
\int d^3 \bfx\, V^a(\bfx) 
\left[4\pi i^J \Psi_{(JM)a}^{\alpha,k}(\bfx) \right]^*
=
\int d^3 \bfx\, V^a(\bfx) 
\left[4\pi i^J \mathcal{O}^\alpha_a \Psi_{(JM)}^{k}(\bfx) \right]^*
\nn\\
&=&
\int d^3 \bfx\, 
\left[
\left(\mathcal{O}^\alpha_a \right)^\dag
V^a(\bfx) 
\right]
\left[
4\pi i^J 
\Psi_{(JM)}^{k}(\bfx) 
\right]^*,
\label{eqn:ibp}
\ea
where $\mathcal{O}^\alpha_a = \{D_a, K_a, M_a\}$, and the 
hermitian conjugates of each operator are as given in
Eq.~(\ref{eqn:DKM_conjugate}).  Explicit expressions for the
expansion coefficients are  
\ba
V_\JM^L(k)
&=&
\int d^3 \bfx\, 
\left[
4\pi i^J 
\Psi_{(JM)}^{k}(\bfx) 
\right]^*
D_a
V^a(\bfx),
\\
V_\JM^B(k)
&=&
\int d^3 \bfx\, 
\left[
4\pi i^J 
\Psi_{(JM)}^{k}(\bfx) 
\right]^*
\left(-K_a\right)
V^a(\bfx),
\\
V_\JM^E(k)
&=&
\int d^3 \bfx\, 
\left[
4\pi i^J 
\Psi_{(JM)}^{k}(\bfx) 
\right]^*
\left(-M_a+2D_a\right)
V^a(\bfx),
\ea
Likewise the coefficients for the helicity basis are
\be
V_\JM^{\lambda=0}(k) = V_\JM^L(k),\qquad
V_\JM^{\lambda=\pm 1}(k) = \frac{1}{\sqrt{2}}
\left[
V_\JM^E(k) \mp i V_\JM^B
\right].
\ee 
In other words, the expansion coefficients for vector TAM waves 
are the same as the coefficients of the scalar TAM waves for 
following three scalar functions:
\ba
V^L(\bfx) &=& D_a V^a(\bfx) 
= \frac{i}{k}\nabla_a V^a(\bfx) = \frac{i}{k}\nabla\cdot {\bf V}(\bfx),
\\
V^B(\bfx) &=& -K_a V^a(\bfx) = \epsilon_{abc}x^b\nabla^c V^a(\bfx)
=
\left[\bfx\times\nabla\right]\cdot {\bf V}(\bfx),
\\
V^E(\bfx) &=&
(-M_a+2D_a) V^a(\bfx) = 
- \frac{i}{k}\epsilon_{abc}\nabla^b K^c V^a(\bfx)
+2 \frac{i}{k}\nabla_aV^a(\bfx)
=
\frac{i}{k}
\left[
\left\{\nabla\times\left(\bfx\times\nabla\right)\right\}
+
2\nabla
\right]\cdot {\bf V}(\bfx).
\label{eqn:scalarfunctions}
\ea
These scalars may be useful to calculate the theoretical
expectation for TAM-wave coefficients.

Suppose now that $V_a({\bf x})$ is written in terms of its
longitudinal and transverse parts and that these have power
spectra $P_L(k)$ and $P_T(k)$, as defined in
Section~\ref{sec:vectorplanewaves}.  It follows then that
\begin{equation}
     \VEV{ \left[V_{(JM)}^\alpha(k)\right]^* V_{(J'M')}^\beta(k')} =
     P_T(k) \delta_{JJ'} \delta_{MM'} \delta_{\alpha\beta}
     \frac{(2\pi)^3}{k^2} \delta_D(k-k'),
\end{equation}
for $\{\alpha,\beta\} = \{E,B\}$.  Similarly, 
\begin{equation}
     \VEV{ \left[V_{(JM)}^L(k)\right]^* V_{(J'M')}^L(k')} =
     P_L(k) \delta_{JJ'} \delta_{MM'} 
     \frac{(2\pi)^3}{k^2} \delta_D(k-k'),
\end{equation}
for the longitudinal modes.  The projections of a
linearly-polarized transverse-vector plane wave onto the $E$ and
$B$ vector TAM waves have equal amplitudes.  Therefore, the
power spectra for the $E$ and $B$ modes must always be the same
for a realization of a statistically homogeneous random field.

\section{Symmetric Tensor Fields}
\label{sec:tensorfields}

\subsection{Introduction and Plane Waves}

We now consider solutions to the Helmholtz equation,
$\left(\nabla^{2}+k^{2}\right) h_{ab}(\bfx)=0$,
for a symmetric tensor field
$h_{ab}(\bfx)=h_{(ab)}(\bfx) \equiv
\left[h_{ab}(\bfx)+h_{ba}(\bfx) \right]/2$.
The most general such tensor field can be decomposed into a
trace component $h(\bfx)$, a longitudinal component $\xi(\bfx)$, two
vector components $w_a$ (with
$\nabla^a w_a=0$), and two transverse-traceless tensor
components $h_{ab}^{TT}$ (which satisfy 
$\nabla^a h_{ab}^{TT}=0$ and $h^a{}_a=0$), as
\be
     h_{ab} = h
     g_{ab}+\left(\nabla_{a}\nabla_{b}-\frac{1}{3}g_{ab}
     \nabla^{2}\right)\xi+\nabla_{(a}w_{b)}+h_{ab}^{TT}.
\label{eqn:SVTdecomposition}
\ee
Our goal is to derive rank-2 tensor solutions to the Helmholtz
equation, of definite total angular momentum, for these
different components.

We begin, though, by reviewing the Fourier decomposition of the
rank-2 tensor field.  Each Fourier component of the tensor field
can be expanded as $\tilde h_{ab}(\bfk) = \sum_s \varepsilon^s_{ab}(\hatk)
h_s(\bfk)$ in terms of six polarization states
$\varepsilon^s_{ab}(\bfk)$, where $s=\{0,z,x,y,+,\times\}$, for the
trace, longitudinal, two vector, and two transverse-traceless
polarizations, respectively, with amplitudes $h_s(\bfk)$
\cite{Jeong:2012df}.  The polarization tensors satisfy
$\varepsilon^{s\,ab} \varepsilon^{s'}_{ab}=2\delta_{ss'}$.  The trace
polarization tensor is $\varepsilon^0_{ab}\propto \delta_{ab}$, and the
longitudinal is
$\varepsilon^z_{ab} \propto (k^a k^b-k^2 \delta_{ab}/3)k^{-2}$.
The two vector-mode polarization tensors
satisfy $\varepsilon^{x,y}_{ab} \propto k_{(a} w^{x,y}_{b)}$ where
$w^{x,y}_a$ are two orthogonal ($w^{x\,a} w^y_a=0$) and transverse
($k^a w^{x,y}_a=0$) vectors.  The two transverse-traceless
polarization states have $k^a \varepsilon^{+,\times}_{ab}=0$.

The two vector (spin-1) modes $x,y$ can alternatively
be written in terms of a helicity basis by defining two
helicity-1 polarization tensors $\varepsilon^{\pm1}_{ab} =
(\varepsilon^x_{ab} \pm i \varepsilon^y_{ab})/\sqrt{2}$.
Similarly, the two transverse-traceless (spin-2) modes $+,\times$ can
alternatively be written in terms of a helicity basis by
defining two helicity-2 polarization tensors $\varepsilon^{\pm2}_{ab} =
(\varepsilon^+_{ab} \pm i \varepsilon^\times_{ab})/ \sqrt{2}$.

In general relativity, power spectra $P_h(k)$ for gravitational
waves (transverse-traceless tensor fields) $h_{ab}$ are defined,
for example, by
\begin{equation}
     \VEV{h_s(\bfk) h_{s'}(\bfk')} = \delta_{ss^{\prime}} (2\pi)^3
     \delta_D(\bfk-\bfk') \frac{P_h(k)}{4},
\label{def:pk_tensor}
\end{equation}
for $s,s'=\{+,\times\}$, so that
\begin{equation}
\left<
\tilde h_{ab}(\bfk) \tilde h^{ab}(\bfk)
\right> = (2\pi)^3 P_h(k)\delta_{ss'}\delta_D(\bfk-\bfk').
\end{equation}

\subsection{TAM waves}

Our aim now is to find tensor-valued functions $h_{ab}(\bfx)$,
solutions to the tensor Helmholtz equation
for wavenumber $k$, that transform under spatial
rotation as representations of
order $J$. These will be eigenfunctions of total
angular momentum $\mathbf{J}=\mathbf{L}+\mathbf{S}$.  Here the
spin can be either $S=0$ for the trace of $h_{ab}$ or $S=2$ for
the trace-free part.  The expansion of the trace is simply in
terms of scalar TAM waves.  We will therefore focus our
attention in the following on trace-free rank-2 tensors.
Therefore, $\mathbf{S}$ is now the $S=2$ spin associated
with the vector space spanned by a set of basis tensors at each
spatial point.

We start by constructing a rank-2 spherical basis
$t^{\tilde{m}}_{ab}$, for
$\tilde{m}=\pm2,\pm1,0$, that
transforms under rotations as a representation of order 2
by taking direct products of the order-1 spherical
basis,\footnote{Tilded indices like $\tilde{m}$ are reserved for
the order-2 spherical basis.}
\begin{equation}
     t^{\tilde{m}}_{ab}\equiv\sum_{\bar{m}_{1}\bar{m}_{2}}
     \langle1\bar{m}_{1}1\bar{m_{2}}|2 \tilde{m}\rangle
     e^{\bar{m}_{1}}_a e^{\bar{m}_{2}}_b.
\label{def:spherical_basis_tensor}
\end{equation}
Using orthonormality of Clebsch-Gordan coefficients, these are
normalized to $\left(t^{\tilde{m}_{1}}\right)^{ab}
t^{\tilde{m}_{2}\,*}_{\, ab}  = \delta_{\tilde{m}_{1}
\tilde{m}_{2}}$.

\subsubsection{The orbital-angular-momentum basis}

We begin by expanding the five components of the rank-2
traceless tensor in terms of five tensor TAM waves of definite
orbital-angular-momentum-squared ${\bf L}^2$ for each
total angular momentum $JM$, as
\begin{equation}
     \Psi_{(JM)ab}^{l,k}(\bfx) \equiv
     j_{l}\left(kr\right) Y_{(JM)ab}^{l,k}(\hatn) =
     \sum_{\tilde{m}m}\langle2\tilde{m}lm|JM\rangle
     j_{l}\left(kr\right)Y_{(lm)}(\hatn) t^{\tilde{m}}_{ab},\qquad
     l=J-2,J-1,J,J+1,J+2,
\label{def:Psiabl}
\end{equation}
an equation that also defines the OAM tensor spherical
harmonics $Y_{(JM)ab}^{l,k}(\hatn)$.  These OAM tensor spherical
harmonics of fixed orbital angular momentum satisfy the
orthonormality relation,
\begin{equation}
     \int\, d^2\hatn\, {Y}_{(JM)}^{l\,\,ab}(\hatn)
     Y_{(J'M')ab}^{l'\,*}(\hatn) =
     \delta_{ll'}\delta_{JJ'}\delta_{MM'}.
\end{equation}
The demonstration that the $\Psi_{(JM)ab}^{l,k}(\bfx)$
constitute a complete basis for traceless symmetric tensors on
$\mathbb{R}^3$, and that $Y_{(JM)ab}^l(\hatn)$ constitute a
complete basis for three-dimensional traceless tensors on
$\mathbb{S}^2$,
are straightforward and similar to the analogous proofs for
vector harmonics presented in Section \ref{sec:OAMvector}.

\subsubsection{The longitudinal/vector/transverse-traceless basis}

We now proceed to write the five traceless tensor harmonics for
each $JM$ in terms of a longitudinal ($L$) component, two
vector components ($VE$ and $VB$), and two transverse-traceless
components ($TE$ and $TB$).
In Appendix \ref{appendix:tensordivergence} we derive the
divergence of the tensor spherical waves of fixed orbital
angular momentum in terms of vector spherical waves to be
\begin{equation}
     \frac{1}{k}\nabla^a \Psi_{(JM)ab}^{l,k}(\bfx)=
     -
     \begin{cases}
     \sqrt{\frac{J-1}{2J-1}} \Psi^{k,J-1}_{(JM)b}(\bfx), & \qquad
     l=J-2,\\ \sqrt{\frac{J-1}{2 \left(2J+1\right)}}
     \Psi_{(JM)b}^{k,J}(\bfx), & \qquad l=J-1,\\
     \sqrt{\frac{\left(J+1\right) \left(2J+3\right)}{6
     \left(2J-1\right) \left(2J+1\right)}}
     \Psi^{k,J-1}_{(JM)b}(\bfx) +
     \sqrt{\frac{J\left(2J-1\right)}{6\left(2J+1\right)
     \left(2J+3\right)}} \Psi^{k,J+1}_{(JM)b}(\bfx),
     & \qquad l=J,\\ 
     \sqrt{\frac{J+2}{2\left(2J+1\right)}}
     \Psi^{k,J}_{(JM)b}(\bfx),  & \qquad l=J+1,\\
     \sqrt{\frac{J+2}{2J+3}} \Psi^{k,J+1}_{(JM)b}(\bfx), & \qquad
     l=J+2.
\end{cases}
\end{equation}
Note that the divergence of a tensor of fixed total angular
momentum $JM$ yields a vector of the same $JM$, since we have
acted with $\nabla_a$, an irreducible-vector operator.

{\it The transverse-traceless modes.}
We can, from these results, immediately construct two linear
combinations of $\Psi^{k,l}_{(JM)ab}(\bfx)$, of different parity, with
vanishing divergence,
\begin{align}
     &\Psi_{(JM)ab}^{TE}(\bfx) \equiv \left(\frac{\left(J+1\right)
     \left(J+2\right)}{2\left(2J-1\right)\left(2J+1\right)}\right)^{1/2}
     \Psi^{J-2}_{(JM)ab}(\bfx)-\left(\frac{3\left(J-1\right)
     \left(J+2\right)}{\left(2J-1\right) \left(2J+3\right)}
     \right)^{1/2} \Psi^J_{(JM)ab}(\bfx) \nn \\
     & \qquad \qquad \qquad +
     \left(\frac{J\left(J-1\right)}{2\left(2J + 1\right)
     \left(2J+3\right)}\right)^{1/2} \Psi^{J+2}_{(JM)ab}(\bfx),\nn\\ 
     &\Psi^{TB}_{(JM)ab}(\bfx) \equiv
     \left(\frac{J+2}{2J+1}\right)^{1/2}
     \Psi^{J-1}_{(JM)ab}(\bfx)-\left(\frac{J-1}{2J+1}\right)^{1/2}
     \Psi^{J+1}_{(JM)ab}(\bfx).
\end{align}
These two spherical waves form a basis for the
transverse-traceless (TT) part of the tensor.  We label them $E$
and $B$ according to their parity, $(-1)^{J}$ or $(-1)^{J+1}$,
respectively.

{\it The vector modes.}
The divergence of the vector component of the rank-2
tensor yields a divergence-free vector field.  Recalling that
the vector harmonics $i\Psi^{J}_{(JM)a}(\bfx) = \Psi^{B}_{(JM)a}(\bfx)$ are
divergence-free, we can construct one vector mode of
the tensor field by taking the other orthogonal linear
combination of $\Psi^{J-1}_{(JM)ab}(\bfx)$ and
$\Psi^{J+1}_{(JM)ab}(\bfx)$. Likewise, the other vector
mode should have a divergence that is proportional to
$\Psi^E_{(JM)a}(\bfx)$, but it should also be orthogonal to the
transverse-traceless modes we already obtained. After some
algebra, we find the two vector modes of the tensor to be
\begin{align}
     &\Psi^{VB}_{(JM)ab}(\bfx) = \left(\frac{J-1}{2J+1}\right)^{1/2}
     \Psi^{J-1}_{(JM)ab}(\bfx) + \left(\frac{J+2}{2J+1}\right)^{1/2}
     \Psi^{J+1}_{(JM)ab}(\bfx), \nn\\
     &\Psi^{VE}_{(JM)ab}(\bfx) =
     \left(\frac{2\left(J-1\right)\left(J+1\right)}
     {\left(2J-1\right)\left(2J+1\right)}\right)^{1/2} 
     \Psi^{J-2}_{(JM)ab}(\bfx) + \left(\frac{3}{\left(2J-1\right)
     \left(2J+3\right)}\right)^{1/2} \Psi^J_{(JM)ab}(\bfx) \nn
     \\
     & \qquad \qquad \qquad -
     \left(\frac{2J\left(J+2\right)} {\left(2J+1\right) 
     \left(2J+3\right)}\right)^{1/2} \Psi^{J+2}_{(JM)ab}(\bfx).
\end{align}
These form a basis for the vector part of the tensor
field.  The divergences of these basis functions are,
\begin{equation}
     \frac{1}{k} \nabla^a
     \Psi^{VB}_{(JM)ab}(\bfx)= \frac{i}{\sqrt{2}}
     \Psi^{B}_{(JM)b}(\bfx),\qquad
      \frac{1}{k} \nabla^a \Psi^{VE}_{(JM)ab}(\bfx)=
     \frac{i}{\sqrt{2}} \Psi^E_{(JM)b}(\bfx).
\end{equation}
It then follows that the vector TAM waves of
the tensor field can be obtained from the transverse-vector
spherical waves through,
\begin{equation}
     \Psi^{VB}_{(JM)ab}(\bfx)=-\frac{i}{k\sqrt{2} }
     \left( \nabla_a \Psi^B_{(JM)b}(\bfx) + \nabla_b \Psi^B_{(JM)a}(\bfx)
     \right), \qquad
     \Psi^{VE}_{(JM)ab}(\bfx)=-\frac{i}{k\sqrt{2} }
     \left( \nabla_a \Psi^E_{(JM)b}(\bfx) + \nabla_b \Psi^E_{(JM)a}(\bfx)
     \right).
\label{eqn:vectordivs}
\end{equation}

{\it The longitudinal mode.}
The last orthogonal linear combination of the
orbital-angular-momentum states,
\begin{align}
     \Psi^L_{(JM)ab}(\bfx) = &
     \left(\frac{3\left(J-1\right)J}{2\left(2J-1\right)
     \left(2J+1\right)}\right)^{1/2} \Psi^{J-2}_{(JM)ab}(\bfx) +
     \left(\frac{J\left(J+1\right)} {\left(2J-1\right)
     \left(2J+3\right)}\right)^{1/2} \Psi^J_{(JM)ab}(\bfx) \nn
     \\
     & \qquad \qquad \qquad +
     \left(\frac{3\left(J+1\right)\left(J+2\right)}
     {2\left(2J+1\right)\left(2J+3\right)}\right)^{1/2}
     \Psi^{J+2}_{(JM)ab}(\bfx),
\end{align}
decries the longitudinal component.  To check, we find its
divergence to be
\be
     \frac{1}{k}\nabla^a \Psi^L_{(JM)ab}(\bfx) =
     i\sqrt{\frac{2}{3}} \Psi^L_{(JM)b}(\bfx).
\ee
It implies that the longitudinal mode is the only one that has
non-vanishing double divergence,
\be
     \frac{1}{k^{2}}\nabla^a \nabla^b \Psi^L_{(JM)ab} (\bfx)=
     \sqrt{\frac{2}{3}} \Psi_{(JM)}(\bfx),
\ee
in terms of the scalar spherical wave $\Psi_{(JM)}$.  Using this
result, it further follows that,
\be
\label{eq:tensor-longitudinal}
     \Psi^L_{(JM)ab}(\bfx) = \frac{1}{k^2}
     \sqrt{\frac{3}{2}}
     \left(\nabla_a\nabla_b-\frac{1}{3} g_{ab}
     \nabla^{2}\right)\Psi_{(JM)}(\bfx).
\ee

\subsubsection{The longitudinal/vector/transverse-traceless
basis in terms of derivative operators}
\label{sec:tensoroperators}

As seen in Section \ref{sec:vectoroperators}, the $L$, $B$, and $E$
vector waves can be written by applying the vector operators
$D_a$, $K_a$, and $M_a$, respectively, to scalar TAM
waves. Likewise, we have just seen in
Eq.~(\ref{eq:tensor-longitudinal}) that the TAM wave for
the longitudinal component of the tensor field can be written by
applying a derivative operator $\nabla_a
\nabla_b-(1/3)g_{ab} \nabla^2$ to the scalar spherical
wave.  We have also seen in Eq.~(\ref{eqn:vectordivs}) that TAM
waves for the vector
components of the tensor field can be written by taking a
symmetrized gradient of the transverse-vector spherical
harmonics; i.e., by applying the operators $\nabla_{(a} K_{b)}$
and $\nabla_{(a} M_{b)}$, to scalar spherical waves.  We now
present an operator approach so that we have a complete
treatment of symmetric traceless tensors in terms of tensor
differential operators, including the two transverse-traceless
tensor modes.

Using the three vector operators $D_a$, $K_a$, and $M_a$ we have
proposed as basic building blocks, we construct five tensor
operators
\ba
T_{ab}^{L}  &=& -D_{a}D_{b}+\frac{1}{3} g_{ab}, \qquad
T_{ab}^{VB}  = D_{(a}K_{b)}, \qquad
T_{ab}^{VE}  = D_{(a}M_{b)}, \nn \\
T_{ab}^{TB}  &=& K_{(a}M_{b)}+M_{(a}K_{b)}+2D_{(a}K_{b)}, \qquad
T_{ab}^{TE}  = M_{(a}M_{b)}-K_{(a}K_{b)}+2D_{(a}M_{b)}.
\label{def:Tab}
\ea
These are irreducible tensors since they are symmetric and
traceless, and they commute with $\nabla^2$. Therefore, when
acting on scalar TAM waves $\Psi^{k}_{(JM)}$, they generate
symmetric tensor TAM waves that solve the tensor Helmholtz
equation and have the same total angular momentum $JM$. We have
seen that $T_{ab}^{L}$ generates longitudinal mode, and
$T_{ab}^{VB}$ and $T_{ab}^{VE}$ generate $B$ and $E$ vector
modes, respectively. It is straightforward to show that
$D^a T_{ab}^{TB}=D^a T_{ab}^{TE}=0$, so they generate
transverse-traceless tensor modes. The parity-odd
$T_{ab}^{TB}$ generates the $B$ mode
$\Psi_{\left(JM\right)ab}^{TB,k}(\bfx)\propto
T_{ab}^{TB}\Psi_{\left(JM\right)}^{k}(\bfx)$, while the parity-even
$T_{ab}^{TE}$ generates the $E$ mode,
$\Psi_{\left(JM\right)ab}^{TE,k}(\bfx)\propto
T_{ab}^{TE}\Psi_{\left(JM\right)}^{k}(\bfx)$. The five operators
generate five linearly independent tensor modes, because they
are orthogonal according to
\be
    \left(T_{ab}^{\alpha^{\prime}}\right)^{\dagger}
    T^{\alpha,ab}=0, \quad
    \mathrm{if\,\alpha\neq\alpha^{\prime}},
    \qquad\mathrm{for\,}\qquad \alpha,\alpha^{\prime}=L,VB,VE,TB,TE.
\ee
To normalize the tensor spherical waves, we calculate the norms
of those five tensor operators to be
\begin{align}
     &\left(T_{ab}^{L}\right)^{\dagger}T^{L,ab} =
     \frac{2}{3},\qquad\left(T_{ab}^{VB}\right)^{\dagger}T^{VB,ab}
     =
     \left(T_{ab}^{VE}\right)^{\dagger}T^{VE,ab} =
     \frac{J\left(J+1\right)}{2},\nn\\  
     &\left(T_{ab}^{TE}\right)^{\dagger} T^{TE,ab} =
     \left(T_{ab}^{TB}\right)^{\dagger}T^{TB,ab} =
     \frac{2(J+2)!}{(J-2)!}.
\end{align}

To summarize, the decomposition of the traceless symmetric
rank-2 tensor into longitudinal, vector, and transverse tensor
modes is
\ba
     \mathrm{longitudinal}:\; \Psi_{\left(JM\right)ab}^{L}(\bfx) &=&
     \sqrt{\frac{3}{2}}T_{ab}^{L}\Psi_{\left(JM\right)}(\bfx) =
     \left(\frac{3\left(J-1\right)J}{2\left(2J-1\right)
     \left(2J+1\right)}\right)^{\frac{1}{2}}
     \Psi_{\left(JM\right)ab}^{J-2}(\bfx) \nn\\
      & & +\left(\frac{J\left(J+1\right)}{\left(2J-1\right)\left(2J+3\right)}\right)^{\frac{1}{2}}\Psi_{\left(JM\right)ab}^{J}(\bfx)+\left(\frac{3\left(J+1\right)\left(J+2\right)}{2\left(2J+1\right)\left(2J+3\right)}\right)^{\frac{1}{2}}\Psi_{\left(JM\right)ab}^{J+2}(\bfx),\nn
\ea
\ba
\mathrm{vector\,B\,mode}:\;\Psi_{\left(JM\right)ab}^{VB}(\bfx)&=&-\sqrt{\frac{2}{J\left(J+1\right)}}T_{ab}^{VB}\Psi_{\left(JM\right)}(\bfx)\nn\\ 
  &=&\left(\frac{J-1}{2J+1}\right)^{\frac{1}{2}}\Psi_{\left(JM\right)ab}^{J-1}(\bfx)+\left(\frac{J+2}{2J+1}\right)^{\frac{1}{2}}\Psi_{\left(JM\right)ab}^{J+1}(\bfx),\nn
\ea
\ba
\mathrm{vector\, E\,mode}:\;\Psi_{\left(JM\right)ab}^{VE}(\bfx)
& = & -\sqrt{\frac{2}{J\left(J+1\right)}}T_{ab}^{VE}\Psi_{\left(JM\right)}(\bfx)=\left(\frac{2\left(J-1\right)\left(J+1\right)}{\left(2J-1\right)\left(2J+1\right)}\right)^{\frac{1}{2}}\Psi_{\left(JM\right)ab}^{J-2}(\bfx)\nn\\
& & +\left(\frac{3}{\left(2J-1\right)\left(2J+3\right)}\right)^{\frac{1}{2}}\Psi_{\left(JM\right)ab}^{J}(\bfx)-\left(\frac{2J\left(J+2\right)}{\left(2J+1\right)\left(2J+3\right)}\right)^{\frac{1}{2}}\Psi_{\left(JM\right)ab}^{J+2}(\bfx),\nn
\ea
\ba
\mathrm{transverse\, tensor\,
B\,mode}:\;\Psi_{\left(JM\right)ab}^{TB}(\bfx)&=& -
\sqrt{\frac{(J-2)!}{2 (J+2)!}} 
T_{ab}^{TB}\Psi_{\left(JM\right)}(\bfx)\nn\\
&=&\left(\frac{J+2}{2J+1}\right)^{\frac{1}{2}}\Psi_{\left(JM\right)ab}^{J-1}(\bfx)-\left(\frac{J-1}{2J+1}\right)^{\frac{1}{2}}\Psi_{\left(JM\right)ab}^{J+1}(\bfx),\nn
\ea
\ba
\mathrm{transverse\, tensor\,
E\,mode}:\;\Psi_{\left(JM\right)ab}^{TE}(\bfx)&=&
-\sqrt{\frac{(J-2)!}{2 (J+2)!}} 
T_{ab}^{TE}\Psi_{\left(JM\right)}(\bfx)=\left(\frac{\left(J+1\right)\left(J+2\right)}{2\left(2J-1\right)\left(2J+1\right)}\right)^{\frac{1}{2}}\Psi_{\left(JM\right)ab}^{J-2}(\bfx)\nn\\
& & -\left(\frac{3\left(J-1\right)\left(J+2\right)}{\left(2J-1\right)\left(2J+3\right)}\right)^{\frac{1}{2}}\Psi_{\left(JM\right)ab}^{J}(\bfx)+\left(\frac{J\left(J-1\right)}{2\left(2J+1\right)\left(2J+3\right)}\right)^{\frac{1}{2}}\Psi_{\left(JM\right)ab}^{J+2}(\bfx).\nn\\
\label{def:Psiabalpha}
\ea
The normalizations are chosen such that
\begin{equation}
     16\pi^2 \int\ d^3x\,
     \left[\Psi^{\alpha,k}_{(JM)ab}(\bfx)\right]^*
      \Psi^{\beta,k\,\,\,ab}_{(J'M')}(\bfx) =
     \delta_{\alpha\beta} \delta_{JJ'}\delta_{MM'} \frac{(2\pi)^3}{k^2}
     \delta_D(k-k'),
\end{equation}
where $\{\alpha,\beta\}=\{L,VB,VE,TB,TE\}$.  Again, the five
modes are orthogonal as tensor wave functions in the Hilbert
space, but their tensor values at any given point are {\it not}
necessarily orthogonal, as we will see below. This orthogonality
does hold asymptotically in the far-field limit
$kr\rightarrow\infty$.

\subsubsection{Summary and helicity basis}

So far we have constructed two sets of TAM-wave bases for
symmetric traceless tensors. The OAM basis $\Psi^{l}_{(JM)ab}(\bfx)$,
where $l=J-2,J-1,J,J+1,J+2$ for each $JM$, are eigenstates of
the square of orbital angular momentum $\mathbf{L}^2$. We have
also defined a second basis $\Psi^{\alpha}_{(JM)ab}(\bfx)$ in terms of
a longitudinal mode $\alpha=L$, two vector modes $\alpha=VE,VB$,
and two transverse-traceless tensor modes $\alpha=TE,TB$, for
each $JM$. We can furthermore
construct a helicity basis $\Psi^{\lambda}_{(JM)ab}(\bfx)$, denoted
by helicity $\lambda=\pm2,\pm1,0$, through
\ba
    && \Psi_{\left(JM\right)ab}^{\pm2}(\bfx)=\frac{1}{\sqrt{2}}
     \left(\Psi_{\left(JM\right)ab}^{TE}(\bfx)\pm
     i\Psi_{\left(JM\right)ab}^{TB}(\bfx)\right), \nn\\
    && \Psi_{\left(JM\right)ab}^{\pm1}(\bfx)=\frac{1}{\sqrt{2}}
     \left(\Psi_{\left(JM\right)ab}^{VE}(\bfx)
     \pm i\Psi_{\left(JM\right)ab}^{VB}(\bfx)\right),\nn \\
    && \Psi_{\left(JM\right)ab}^{0}(\bfx)=\Psi_{\left(JM\right)ab}^{L}(\bfx).
\label{def:Psiablambda}
\ea
These are eigenstates of the helicity operator
$H=\mathbf{S}\cdot\mathbf{\hat{p}}$, but for the tensor field,
$\left(S_e\right)_{ab,cd}\equiv i\epsilon_{aec}g_{bd}+ig_{ac}
\epsilon_{bed}$, and $\hat{p}_a=-(i/k)\nabla_a$. 

The orbital basis $\Psi^{l}_{(JM)ab}(\bfx)$, for $l=J-2,J-1,J,J+2,J+2$,
the longitudinal/vector/transverse-traceless basis
$\Psi^{\alpha}_{(JM)ab}(\bfx)$, for $\alpha=L,VE,VB,TE,TB$, and the
helicity basis $\Psi^{\lambda}_{(JM)ab}(\bfx)$, for
$\lambda=0,+1,-1,+2,-2$, are related by unitary transformations,
\begin{align}
&U_{\alpha l}^{J}=\left(\begin{array}{ccccc}
\sqrt{\frac{3\left(J-1\right)J}{2\left(2J-1\right)\left(2J+1\right)}} & 0 & \sqrt{\frac{J\left(J+1\right)}{\left(2J-1\right)\left(2J+3\right)}} & 0 & \sqrt{\frac{3\left(J+1\right)\left(J+2\right)}{2\left(2J+1\right)\left(2J+3\right)}}\\
\sqrt{\frac{2\left(J-1\right)\left(J+1\right)}{\left(2J-1\right)\left(2J+1\right)}} & 0 & \sqrt{\frac{3}{\left(2J-1\right)\left(2J+3\right)}} & 0 & -\sqrt{\frac{2J\left(J+2\right)}{\left(2J+1\right)\left(2J+3\right)}}\\
0 & \sqrt{\frac{J-1}{2J+1}} & 0 & \sqrt{\frac{J+2}{2J+1}} & 0\\
\sqrt{\frac{\left(J+1\right)\left(J+2\right)}{2\left(2J-1\right)\left(2J+1\right)}} & 0 & -\sqrt{\frac{3\left(J-1\right)\left(J+2\right)}{\left(2J-1\right)\left(2J+3\right)}} & 0 & \sqrt{\frac{J\left(J-1\right)}{2\left(2J+1\right)\left(2J+3\right)}}\\
0 & \sqrt{\frac{J+2}{2J+1}} & 0 & -\sqrt{\frac{J-1}{2J+1}} & 0
\end{array}\right),\nn\\
&U_{\lambda\alpha}^{J}=\left(\begin{array}{ccccc}
1 & 0 & 0 & 0 & 0\\
0 & i/\sqrt{2} & 1/\sqrt{2} & 0 & 0\\
0 & -i/\sqrt{2} & 1/\sqrt{2} & 0 & 0\\
0 & 0 & 0 & i/\sqrt{2} & 1/\sqrt{2}\\
0 & 0 & 0 & -i/\sqrt{2} & 1/\sqrt{2}
\end{array}\right),\nn\\
&U_{\lambda l}^{J}=\left(\begin{array}{ccccc}
\sqrt{\frac{3\left(J-1\right)J}{2\left(2J-1\right)\left(2J+1\right)}} & 0 & \sqrt{\frac{J\left(J+1\right)}{\left(2J-1\right)\left(2J+3\right)}} & 0 & \sqrt{\frac{3\left(J+1\right)\left(J+2\right)}{2\left(2J+1\right)\left(2J+3\right)}}\\
\sqrt{\frac{\left(J-1\right)\left(J+1\right)}{\left(2J-1\right)\left(2J+1\right)}} & i\sqrt{\frac{J-1}{2\left(2J+1\right)}} & \sqrt{\frac{3}{2\left(2J-1\right)\left(2J+3\right)}} & i\sqrt{\frac{J+2}{2\left(2J+1\right)}} & -\sqrt{\frac{J\left(J+2\right)}{\left(2J+1\right)\left(2J+3\right)}}\\
\sqrt{\frac{\left(J-1\right)\left(J+1\right)}{\left(2J-1\right)\left(2J+1\right)}} & -i\sqrt{\frac{J-1}{2\left(2J+1\right)}} & \sqrt{\frac{3}{2\left(2J-1\right)\left(2J+3\right)}} & -i\sqrt{\frac{J+2}{2\left(2J+1\right)}} & -\sqrt{\frac{J\left(J+2\right)}{\left(2J+1\right)\left(2J+3\right)}}\\
\frac12\sqrt{\frac{\left(J+1\right)\left(J+2\right)}{\left(2J-1\right)\left(2J+1\right)}} & i\sqrt{\frac{J+2}{2\left(2J+1\right)}} & -\sqrt{\frac{3\left(J-1\right)\left(J+2\right)}{2\left(2J-1\right)\left(2J+3\right)}} & -i\sqrt{\frac{J-1}{2\left(2J+1\right)}} & \frac12\sqrt{\frac{J\left(J-1\right)}{\left(2J+1\right)\left(2J+3\right)}}\\
\frac12\sqrt{\frac{\left(J+1\right)\left(J+2\right)}{\left(2J-1\right)\left(2J+1\right)}} & -i\sqrt{\frac{J+2}{2\left(2J+1\right)}} & -\sqrt{\frac{3\left(J-1\right)\left(J+2\right)}{2\left(2J-1\right)\left(2J+3\right)}} & i\sqrt{\frac{J-1}{2\left(2J+1\right)}} & \frac12\sqrt{\frac{J\left(J-1\right)}{\left(2J+1\right)\left(2J+3\right)}}
\end{array}\right).
\label{eqn:tensortransformations}
\end{align}

\subsubsection{Projection onto Tensor Spherical Harmonics}

We now describe the projection of three-dimensional traceless
tensor TAM waves onto the two-sphere.  We will begin by
reviewing the projection onto the familiar $E/B$ tensor
spherical harmonics \cite{Kamionkowski:1996ks} in the
$\hat\theta$-$\hat\phi$ space perpendicular to $\hatn$.  We will
then generalize these two basis tensor spherical harmonics to
include three more that will constitute a complete orthonormal
basis for three-dimensional traceless tensors that live on the
two-sphere, parametrized by $\hatn$.

The usual $E/B$ tensor spherical harmonics are defined by
\begin{align}
     &Y_{(JM)AB}^{E}(\hatn) =
     \sqrt{\frac{2}{J\left(J+1\right)\left(J-1\right)\left(J+2\right)}}
     \left(-\nabla_{A}\nabla_{B}+\frac{1}{2}g_{AB}\nabla^{C}\nabla_{C}\right)
     Y_{(JM)}(\hatn),\nn\\ 
     &Y_{(JM)AB}^{B}(\hatn) =
     \sqrt{\frac{1}{2J\left(J+1\right)\left(J-1\right)\left(J+2\right)}}
     \left(\epsilon_{B}{}^{C}\nabla_{C}\nabla_{A}+\epsilon_{A}{}^{C}
     \nabla_{C}\nabla_{B}\right)Y_{(JM)}(\hatn),
\label{eqn:EBtensor}
\end{align}
where here $\{A,B\}=\{\theta,\phi\}$, and $\nabla_A$ is a
covariant derivative on the two-sphere, with metric $g_{AB} =
{\rm diag}(1,\sin^2\theta)$ and antisymmetric tensor $\epsilon_{AB}$.

The operator approach developed in Section
\ref{sec:vectorprojection} for vector spherical harmonics can be
generalized to tensor spherical harmonics. Recall that the operators
$\{N_a,K_a,M_{\perp a}\}$ satisfy the same algebra as the operators
$\{D_a,K_a,M_a\}$ do. Therefore, following the same line of
reasoning we construct five tensor operators,
\ba
W_{ab}^{L} & = & -N_{a}N_{b}+\frac{1}{3} g_{ab}, \qquad
W_{ab}^{VB}  =N_{(a}K_{b)}, \qquad
W_{ab}^{VE}  =N_{(a}M_{\perp b)}, \nn \\
W_{ab}^{TB} & =& K_{(a}M_{\perp b)}+M_{\perp(a}K_{b)}+2N_{(a}K_{b)},\qquad
W_{ab}^{TE}  =M_{\perp(a}M_{\perp b)}-K_{(a}K_{b)}+2N_{(a}M_{\perp b)},
\label{def:Wab}
\ea
for tensor spherical harmonics.  These symmetric and traceless
operators conserve total angular momentum, since they are
irreducible tensors. The last two operators are perpendicular to
the radial direction,
$\hat{n}^aW_{ab}^{TE}=\hat{n}^aW_{ab}^{TB}=0$, so they must
generate the $E/B$ tensor spherical harmonics,
$\hat{n}^aW_{ab}^{TE}Y_{(JM)}(\hatn)\propto Y^{TE}_{(JM)ab}(\hatn)$ and
$\hat{n}^aW_{ab}^{TB}Y_{(JM)}(\hatn)\propto Y^{TB}_{(JM)ab}(\hatn)$,
respectively. Here $Y^{TE}_{(JM)ab}(\hatn)$ and $Y^{TB}_{(JM)ab}(\hatn)$
correspond to the E/B tensor harmonics
$Y^{E}_{(JM)AB}(\hatn),Y^{B}_{(JM)AB}(\hatn)$, but under a three-dimensional
orthonormal basis, denoted by lower-case indices
$a,b,\ldots$. In addition, the $W_{ab}^{VE}$ and $W_{ab}^{VB}$,
when acting on ordinary spherical harmonics $Y_{(JM)}(\hatn)$,
generate $VE/VB$ tensor spherical harmonics
$Y^{VE}_{(JM)ab}(\hatn)$ and $Y^{VB}_{(JM)ab}(\hatn)$,
respectively, with components in both the tangential plane and
normal. Similarly, $W^{L}_{ab}$ generates the longitudinal 
tensor spherical harmonics $Y^{L}_{(JM)ab}(\hatn)$. For the tensor
spherical harmonics, the terms ``longitudinal,'' ``vector,'' and
``transverse,'' refer to the nature of the tensor components
with respect to the normal $\hat n_a$.  To be more
specific, we define
\ba
     Y_{(JM)ab}^{L} (\hatn)&
     =&\sqrt{\frac{3}{2}}W_{ab}^{L}Y_{(JM)}(\hatn), \nn\\
     Y_{(JM)ab}^{VE} (\hatn)
     &=&-\sqrt{\frac{2}{J(J+1)}}W_{ab}^{VE}Y_{(JM)}(\hatn), \qquad
     Y_{(JM)ab}^{VB}(\hatn) 
     =-\sqrt{\frac{2}{J(J+1)}}W_{ab}^{VB}Y_{(JM)}(\hatn),\nn
     \\
     Y_{(JM)ab}^{TE}(\hatn) &
     =&- \sqrt{\frac{(J-2)!}{2 (J+2)!}} 
     W_{ab}^{TE}Y_{(JM)}(\hatn), \qquad
     Y_{(JM)ab}^{TB}(\hatn) 
     =- \sqrt{\frac{(J-2)!}{2 (J+2)!}} W_{ab}^{TB}Y_{(JM)}(\hatn).
\ea
In terms of the OAM tensor spherical harmonics
$Y^l_{(JM)ab}(\hatn)$, for $l=J,J\pm1,J\pm2$, the
longitudinal/vector/transverse-traceless basis
$Y^{\alpha}_{(JM)ab}(\hatn),\alpha=L,VE,VB,TE,TB$,
are found to be
\ba
     \mathrm{longitudinal}:\quad
     Y_{(JM)ab}^{L}(\hatn)&=&-\left(\frac{3J\left(J-1\right)}
     {2\left(2J-1\right)\left(2J+1\right)}\right)^{\frac{1}{2}}
     Y_{(JM)ab}^{J-2}(\hatn)\nn\\
     & &+\left(\frac{J\left(J+1\right)}{\left(2J-1\right)\left(2J+3\right)}\right)^{\frac{1}{2}}Y_{(JM)}^{J}(\hatn)-\left(\frac{3\left(J+1\right)\left(J+2\right)}{2\left(2J+1\right)\left(2J+3\right)}\right)^{\frac{1}{2}}Y_{(JM)}^{J+2}(\hatn),\nn
\ea
\ba
     \mathrm{vector\, E\, mode}:\quad
     Y_{(JM)ab}^{VE}(\hatn) &=
     &-\left(\frac{2\left(J-1\right)\left(J+1\right)}{\left(2J-1\right)\left(2J+1\right)}\right)^{\frac{1}{2}}Y_{(JM)ab}^{J-2}(\hatn)\nn\\ 
& &+\left(\frac{3}{\left(2J-1\right)\left(2J+3\right)}\right)^{\frac{1}{2}}Y_{(JM)ab}^{J}(\hatn)+\left(\frac{2J\left(J+2\right)}{\left(2J+1\right)\left(2J+3\right)}\right)^{\frac{1}{2}}Y_{(JM)ab}^{J+2}(\hatn),\nn
\ea
\be
  \mathrm{vector\, B\, mode}:\quad Y_{(JM)ab}^{VB}(\hatn) =
     i\left(\frac{J-1}{2J+1}\right)^{\frac{1}{2}}Y_{(JM)ab}^{J-1}(\hatn)
     -
     i\left(\frac{J+2}{2J+1}\right)^{\frac{1}{2}}Y_{(JM)ab}^{J+1}(\hatn),\nn
\ee
\ba
     \mathrm{transverse\, tensor\, E\, mode}: \quad
     Y_{(JM)ab}^{TE}(\hatn) &= & -
     \left(\frac{\left(J+1\right)\left(J+2\right)}{2\left(2J-1\right)
     \left(2J+1\right)}\right)^{\frac{1}{2}}
     Y_{(JM)ab}^{J-2}(\hatn)\nn\\
     & &-\left(\frac{3\left(J-1\right)\left(J+2\right)}
     {\left(2J-1\right)\left(2J+3\right)}\right)^{\frac{1}{2}}Y_{(JM)ab}^{J}(\hatn)
     - \left(\frac{J\left(J-1\right)}{2\left(2J+1\right)
     \left(2J+3\right)}\right)^{\frac{1}{2}}Y_{(JM)ab}^{J+2}(\hatn),\nn
\ea
\be
     \mathrm{transverse\, tensor\, B\, mode}: \quad
     Y_{(JM)ab}^{TB}(\hatn) =
     i\left(\frac{J+2}{2J+1}\right)^{\frac{1}{2}}
     Y_{(JM)ab}^{J-1}(\hatn)+i\left(\frac{J-1}{2J+1}\right)^{\frac{1}{2}}
     Y_{(JM)ab}^{J+1}(\hatn).
\label{def:Ylmabl}
\ee
These are normalized to
\begin{equation}
     \int \, d^2\hatn \, \left[Y^{\alpha,ab}_{(JM)} \left( \hatn \right)\right]^*
     Y^{\beta}_{(J'M')ab} \left( \hatn \right) =\delta_{JJ'} \delta_{MM'} \delta_{\alpha\beta},
\end{equation}
for $\alpha,\beta=\{L,VE,VB,TE,TB\}$.  Note again that
although the transformations between the OAM and
$L/VE/VE/TB/TE$ bases for the tensor spherical harmonics
resemble those for the transformations between the analogous
bases for TAM waves, there are important sign differences.

In terms of the tensor spherical harmonics, the TAM waves
$\Psi^{k,\alpha}_{(JM)ab}(\bfx)$ can be written,
\begin{align}
     \Psi_{(JM)ab}^{k,L}\left(\mathbf{x}\right)=&-\frac{1}{2}\left(j_{J}(kr)+3g_{J}\left(kr\right)\right)Y_{(JM)ab}^{L}\left(\mathbf{\hat{n}}\right)-\sqrt{3J\left(J+1\right)}f_{J}\left(kr\right)Y_{(JM)ab}^{VE}\left(\mathbf{\hat{n}}\right)\nn\\
&-\frac{\sqrt{3J\left(J+1\right)\left(J-1\right)\left(J+2\right)}}{2}\frac{j_{J}\left(kr\right)}{\left(kr\right)^{2}}Y_{(JM)ab}^{TE}\left(\mathbf{\hat{n}}\right),\nn\\
\Psi_{(JM)ab}^{k,VE}\left(\mathbf{x}\right)=&-\sqrt{3J\left(J+1\right)}f_{J}\left(kr\right)Y_{(JM)ab}^{L}\left(\mathbf{\hat{n}}\right)-\left(j_{J}\left(kr\right)+2g_{J}\left(kr\right)+2f_{J}\left(kr\right)\right)Y_{(JM)ab}^{VE}\left(\mathbf{\hat{n}}\right)\nn\\
&-\sqrt{\left(J-1\right)\left(J+2\right)}\left(f_{J}\left(kr\right)+2\frac{j_{J}\left(kr\right)}{\left(kr\right)^{2}}\right)Y_{(JM)ab}^{TE}\left(\mathbf{\hat{n}}\right),\nn\\
\Psi_{(JM)ab}^{k,TE}\left(\mathbf{x}\right)=&-\frac{\sqrt{3J\left(J+1\right)\left(J-1\right)\left(J+2\right)}}{2}\frac{j_{J}\left(kr\right)}{\left(kr\right)^{2}}Y_{(JM)ab}^{L}\left(\mathbf{\hat{n}}\right)-\sqrt{\left(J-1\right)\left(J+2\right)}\left(f_{J}\left(kr\right)+2\frac{j_{J}\left(kr\right)}{\left(kr\right)^{2}}\right)Y_{(JM)ab}^{VE}\left(\mathbf{\hat{n}}\right)\nn\\
&-\frac12\left(-j_{J}(kr)+g_{J}\left(kr\right)+4f_{J}\left(kr\right)+6\frac{j_{J}\left(kr\right)}{\left(kr\right)^{2}}\right)Y_{(JM)ab}^{TE}\left(\mathbf{\hat{n}}\right),\nn\\
\Psi_{(JM)ab}^{k,VB}\left(\mathbf{x}\right)=&-i\left(j_{J}'(kr)-\frac{j_{J}\left(kr\right)}{kr}\right)Y_{(JM)ab}^{VB}\left(\mathbf{\hat{n}}\right)-i\sqrt{\left(J-1\right)\left(J+2\right)}\frac{j_{J}\left(kr\right)}{kr}Y_{(JM)ab}^{TB}\left(\mathbf{\hat{n}}\right),\nn\\
\Psi_{(JM)ab}^{k,TB}\left(\mathbf{x}\right)=&-i\sqrt{\left(J-1\right)\left(J+2\right)}\frac{j_{J}\left(kr\right)}{kr}Y_{(JM)ab}^{VB}\left(\mathbf{\hat{n}}\right)-i\left(j_{J}'(kr)+2\frac{j_{J}\left(kr\right)}{kr}\right)Y_{(JM)ab}^{TB}\left(\mathbf{\hat{n}}\right).
\label{eqn:tensorprojections}
\end{align}
Here we have introduced radial profiles,
\begin{equation}
     f_J(x)\equiv \frac{d}{dx}\frac{j_J(x)}{x},\qquad \text{and}
     \qquad
     g_{J}(x)\equiv-j_{J}(x)-2f_{J}(x)+(J-1)(J+2)\frac{j_J(x)}{x^2}.
\end{equation}
The components proportional to the transverse
$Y^{TE}_{(JM)ab}(\hatn)$ and $Y^{TB}_{(JM)ab}(\hatn)$
harmonics are projections onto the 2-sphere.  These are the
components of principal interest for angular measurements on the
sky.  Eq.~(\ref{eqn:tensorprojections}) shows that the different
tensor TAM waves are not everywhere orthogonal, even though they
are orthonormal, although they do become asymptotically
orthogonal in the $kr \gg 1$ limit.  Both the $VB$ and $TB$
tensor TAM waves have projections onto the $VB$ and $TB$ tensor
spherical harmonics.  The $L$, $VE$, and $TE$ TAM waves have
projections onto the $L$, $VE$, and $TE$ tensor spherical
harmonics.  The phases in our definitions of the tensor
spherical harmonics are chosen so that a rotation of a $TE$
($VB$) mode by $45^\circ$ ($90^\circ$) about $\hatn$ produces a
$TB$ ($VB$) mode.

\subsubsection{Plane-wave expansion for traceless tensor fields}

We now determine the transformation between the tensor
plane-wave basis and the tensor TAM-wave bases. We start with
the OAM basis. The $\Psi^{l,k}_{(JM)ab}$ constitute a complete basis,
and we can write
\begin{equation}
     \hat \varepsilon_{ab}(\bfk) e^{i \bfk \cdot \bfx} = \sum_{lJM} 4\pi i^l
     B_{(JM)}^l(\hatk) \Psi^{l,k}_{(JM)ab}(\bfx)= \sum_{lJM} 4\pi i^l
     B_{(JM)}^l(\hatk) j_l(kr) Y^l_{(JM)ab}(\hatn).
\label{eqn:tensorplanewaves}
\end{equation}
Here $\hat \varepsilon_{ab}(\bfk)$ is a normalized polarization tensor
for the plane wave. We use orthonormality of the
$Y_{(JM)a}^l(\hatn)$ to obtain the expansion coefficients,
\begin{equation}
     B_{(JM)}^l(\hatk) = \hat\varepsilon^{ab} (\bfk)
     Y^{l\,\,*}_{(JM)ab}(\hatk).
\end{equation}
We can similarly expand in terms of $\Psi^{k,\alpha}_{(JM)ab}$,
for $\alpha=L,VE,VB,TE,TB$ modes, or
helicity modes, as
\begin{equation}
     \hat\varepsilon_{ab}(\bfk) e^{i \bfk \cdot \bfx} =
     \sum_{\alpha} \sum_{JM} 4\pi i^{J}
     B_{(JM)}^\alpha(\hatk) \Psi^{k,\alpha}_{(JM)ab}(\bfx), \qquad
     \hat\varepsilon_{ab}(\bfk) e^{i \bfk \cdot \bfx} =
     \sum_{\lambda=0,\pm1,\pm2}\sum_{JM} 4\pi i^J
     B_{(JM)}^\lambda(\hatk) \Psi^{k,\lambda}_{(JM)ab}(\bfx),
\end{equation}
where the expansion coefficients are
\begin{equation}
     B_{(JM)}^\alpha(\hatk) = \hat\varepsilon^{ab} (\bfk)
     Y^{\alpha\,*}_{(JM)ab}(\hatk), \qquad
     B_{(JM)}^\lambda(\hatk) = \hat\varepsilon^{ab} (\bfk)
     Y^{\lambda\,*}_{(JM)ab}(\hatk).
\label{def:Ylmablambda}
\end{equation}
Here, the spin-2 tensor spherical harmonics are defined as
$Y^{\lambda=\pm2}_{(JM)ab} = 2^{-1/2} \left[ Y^{TE}_{(JM)ab}
\pm i Y^{TB}_{(JM)ab} \right]$, $Y^{\lambda=\pm1}_{(JM)ab} = 2^{-1/2} \left[ Y^{VE}_{(JM)ab}
\pm i Y^{VB}_{(JM)ab} \right]$, and $Y^{\lambda=0}_{(JM)ab} = Y^{L}_{(JM)ab}$.
These are related to the spin-2 spherical harmonics
${}_{\lambda} Y_{(JM)}(\hatn)$ \cite{NewmanPenrose,Goldberg} by 
\be
     \hat{\varepsilon}_{\lambda^{\prime}}^{ab}(\mathbf{k})
     Y_{(JM)ab}^{\lambda}(\mathbf{\hat{k}}) =
     {}_{-\lambda}Y_{(JM)}\left(\mathbf{\hat{k}}\right)
     \delta_{\lambda\lambda^{\prime}},\qquad \text{for} \qquad
     \lambda,\lambda^{\prime}=0,\pm1,\pm2.
\ee
Here, $\hat{\varepsilon}^{ab}_{\lambda}(\bfk)$, for
$\lambda=0,\pm1,\pm2$, are the polarization tensors for a
tensor plane wave with wavevector $\bfk$ and helicity
$\lambda$. This equation defines our phase convention for
$\hat{\varepsilon}^{ab}_{\lambda}$. In terms of basis vectors in 
spherical coordinates, these are defined as
\be
\hat{\varepsilon}_{\pm 1}^{ab} =
\frac{1}{\sqrt{2}}
\left[
\hat{\varepsilon}_{\pm1}^a
\hat{n}^b
+
\hat{\varepsilon}_{\pm1}^b
\hat{n}^a
\right], \qquad \hat{\varepsilon}_{\pm 2}^{ab} =
-
\hat{\varepsilon}_{\pm1}^a
\hat{\varepsilon}_{\pm1}^b,
\qquad   
\hat\varepsilon_0^{ab}=
\sqrt{\frac{3}{2}}
\left(\frac13\delta_{ab}- \hat n^a \hat n^b\right),
\label{eq:helicity_polarization_tensor}
\ee
where $\hat{\varepsilon}_0^a$ and $\hat{\varepsilon}_{\pm1}^a$
are defined in  Eq.~(\ref{eq:helicity_polarization_vector}).

\subsubsection{Expansion of tensor fields and power spectra}

An arbitrary symmetric traceless tensor field $h_{ab}(\bfx)$ can be
expanded in the orbital-angular-momentum basis by
\begin{equation}
      h_{ab}({\bf x}) = \sum_{JM} \sum_{l=J,J\pm1,J\pm2} \int
     \frac{k^2 dk}{(2\pi)^3} h_{(JM)}^l(k) 4\pi i^l
     \Psi^{l,k}_{(JM)ab}(\bfx),
\label{eqn:tensorexpansioncoeffs}
\end{equation}
with expansion coefficients,
\begin{equation}
     h_{(JM)}^l(k) = \int d^3 \bfx\, h^{ab}(\bfx) \left[4
     \pi i^l \Psi_{(JM)ab}^{l,k}(\bfx) \right]^*.
\end{equation}
These can also be written as tensor-spherical-harmonic
transforms \cite{inprogress},
\be
     T^{l,k}_{(JM)} = \int \, d^2\hatk\, \tilde T^{ab}(\bfk)
     Y^{l\,\,*}_{(JM)ab}(\hatk),
\ee
of the tensor Fourier amplitudes.
Again, similar relations hold for the $L/VE/VB/TE/TB$ and
helicity bases as well.

The expansion coefficients for the $L/VE/VB/TE/TB$ 
basis can also be re-written, by integrating by parts, as
\ba
h_{(JM)}^\alpha(k) 
&=& 
\int d^3 \bfx\, h^{ab}(\bfx) 
\left[4\pi i^J \Psi_{(JM)ab}^{\alpha,k}(\bfx) \right]^*
=
\int d^3 \bfx\, h^{ab}(\bfx) 
\left[4\pi i^J T^\alpha_{ab} \Psi_{(JM)}^{k}(\bfx) \right]^*
\nn\\
&=&
\int d^3 \bfx\, 
\left[
\left(T^\alpha_{ab} \right)^\dag
h^{ab}(\bfx) 
\right]
\left[
4\pi i^J 
\Psi_{(JM)}^{k}(\bfx) 
\right]^*,
\ea
where $T_{ab}^\alpha$ are the operators defined in Eq.~(\ref{def:Tab}).
Their hermitian conjugates are given by
\ba
\left(T^L_{ab} \right)^\dag &=&
-D_aD_b + \frac13 g_{ab}, \qquad
\left(T^{VE}_{ab} \right)^\dag =
-M_{(a}D_{b)} + 2 D_a D_b,
\qquad \left(T^{VB}_{ab} \right)^\dag =
-K_{(a}D_{b)} ,
\\
\left(T^{TE}_{ab} \right)^\dag &=&
M_{(a}M_{b)} - K_{(a}K_{b)} - 4 M_{(a}D_{b)} - 2 D_{(a}M_{b)} + 8 D_a D_b,
\\
\left(T^{TB}_{ab} \right)^\dag &=&
M_{(a}K_{b)} + K_{(a}M_{b)} - 2D_{(a}K_{b)} - 4K_{(a}D_{b)}.
\ea
We can thus write the expansion coefficients for tensor TAM waves 
as coefficients of the scalar TAM wave for the following scalar
functions:
\ba
h^L(\bfx) &=& 
\left[ -D_aD_b + \frac13 g_{ab}\right]h^{ab}(\bfx)
=
\frac1{k^2}\left[\nabla_a\nabla_b
-\frac{1}{3} g_{ab}\nabla^2 \right] h_{ab}(\bfx),
\\
h^{VE}(\bfx) &=& 
-\frac{1}{k^2}
\left[
\frac12\epsilon_{ac}{}^d\nabla^cK_d\nabla_b
+
\frac12\epsilon_{bc}{}^d\nabla^cK_d\nabla_a
+
2\nabla_a\nabla_b
\right]
h^{ab}(\bfx),
\\
h^{VB}(\bfx) &=&
-\frac{i}{k}
K_{(a}\nabla_{b)}
h^{ab}(\bfx),
\\
h^{TE}(\bfx) &=&
\Biggl[
\frac{1}{k^2}
\Biggl\{
-\frac12
\left(
\epsilon_{ac}{}^d\nabla^c K_d
\epsilon_{be}{}^f\nabla^e K_f
+
\epsilon_{bc}{}^d\nabla^c K_d
\epsilon_{ae}{}^f\nabla^e K_f
\right)
  +2\epsilon_{ac}{}^d\nabla^c K_d\nabla_b
  +2\epsilon_{bc}{}^d\nabla^c K_d\nabla_a
\nn
\\
&&  +\nabla_a \epsilon_{bc}{}^d\nabla^c K_d
  +\nabla_b \epsilon_{ac}{}^d \nabla^c K_d
  -8\nabla_a\nabla_b
\Biggl\}
-K_{(a}K_{b)}
\Biggl]h^{ab}(\bfx),
\\
h^{TB}(\bfx) &=&
-\frac12\frac{i}{k}
\left[
\epsilon_{ac}{}^d \nabla^c K_d K_b
+
\epsilon_{bc}{}^d\nabla^c K_d K_a
+
K_a \epsilon_{bc}{}^d\nabla^c K_d
+
K_b \epsilon_{ac}{}^d \nabla^c K_d
+
2 \nabla_{(a} K_{b)}
+
4 K_{(a} \nabla_{b)}
\right]h^{ab}(\bfx). \nn \\
\label{eqn:tensorscalarfunctions}
\ea
Here, $K_a = \epsilon_{abc}x^b \nabla^c$. 

If $h_{ab}({\bf x})$ is written in terms of
longitudinal/vector/transverse-traceless parts, and if these
have power spectra $P_L(k)$, $P_V(k)$, and $P_T(k)$, then
\begin{align}
     &\VEV{ \left[h_{(JM)}^\alpha(k)\right]^* h_{(J'M')}^\beta(k')} =
     P_T(k) \delta_{JJ'} \delta_{MM'} \delta_{\alpha\beta}
     \frac{(2\pi)^3}{k^2} \delta_D(k-k'), \qquad \alpha,\beta=TE,TB.\\
     &\VEV{ \left[h_{(JM)}^\alpha(k)\right]^* h_{(J'M')}^\beta(k')} =
     P_V(k) \delta_{JJ'} \delta_{MM'} \delta_{\alpha\beta}
     \frac{(2\pi)^3}{k^2} \delta_D(k-k'), \qquad \alpha,\beta=VE,VB.\\
     &\VEV{ \left[h_{(JM)}^L(k)\right]^* h_{(J'M')}^L(k')} =
     P_L(k) \delta_{JJ'} \delta_{MM'} \delta_{\alpha\beta}
     \frac{(2\pi)^3}{k^2} \delta_D(k-k').
\end{align}
For both the vector and transverse-traceless modes, the $E$
modes and $B$ modes have the same power spectra, a consequence
of statistical homogeneity.

\section{Calculation of Lensing Power Spectra}
\label{sec:lensing}

In this Section we provide as an example of the TAM-wave
formalism the calculation of lensing power spectra by density
perturbations and by gravitational waves.  We will reproduce
results from previous work
\cite{Dodelson:2003bv,Cooray:2005hm,Li:2006si,Book:2011na,Namikawa:2011cs},
which were obtained with the Fourier expansion.
For clarity, we only take into account the weak-lensing 
contribution from the deflectors (density perturbations or
gravitational waves) along the line of sight. However, the
measured weak-lensing signal also contains other contributions.
In particular, metric shear,
gravitational-wave effects at the source location
\cite{Dodelson:2003bv}, and tidal alignment dominate the power spectrum
for lensing by gravitational waves \cite{Schmidt:2012}.

Our aim here will be to calculate the lensing deflection field,
\begin{equation}
     \Delta_{a}(\hatn) = \frac{\Pi_{ae}}{\eta_0-\eta}
     \int_{\eta_{0}}^{\eta} \,
     d\eta'\,\left[\hat n_{b}h_{eb} - \frac{1}{2}
     \left(\eta' -
     \eta\right)\hat n_{c}\hat n_{d}\nabla_{e}h_{cd}
     \right]_{\left(\eta',\left(\eta_{0} -
     \eta'\right) \hatn\right)},
\label{eqn:deflection-definition}
\end{equation}
where $\Pi_{ab}\equiv g_{ab}-\hat n_{a} \hat n_{b}$ is the projection
tensor onto the tangential plane.  Thus, the deflection field has no
radial component and can be viewed as a two-dimensional vector
field on the two-sphere.  Here, $h_{ab}(\eta,\bfx)$ is a
(rank-2) tensor metric perturbation evaluated at conformal time
$\eta$.

The deflection field $\Delta_{a}(\hatn)$ on the two-sphere
can be decomposed into gradient and curl components,
\begin{equation}
     \Delta_{a}(\hatn)= 
		M_{\perp a}\varphi(\hatn) + K_a \Omega(\hatn),
\label{eqn:deflectionbreakdown}
\end{equation}
where $M_{\perp a}$ and $K_a$ are the two transverse-vector
operators in Eq.~(\ref{def:NaMaKa}), and $\varphi(\hatn)$ and
$\Omega(\hatn)$ are projected  lensing potentials.
Here we will calculate the contribution
$\Delta_a(\hatn)$ that arises from (a) a 
scalar TAM wave of angular momentum $JM$; 
(b) an $E$ mode transverse-traceless TAM
wave of $JM$; and (c) a $B$ mode transverse-traceless TAM
wave of $JM$.  We will then be able to reproduce the power
spectra $C_J^{\varphi\varphi}$ and $C_J^{\Omega\Omega}$ for density
perturbations and gravitational waves that have been
obtained earlier by considering individual Fourier modes.  These
power spectra are defined by,
\begin{equation}
     \langle\varphi_{(JM)}\varphi_{(J'M')}^{*}\rangle
     =\delta_{JJ'}\delta_{MM'}C_{J}^{\varphi\varphi},
     \qquad
     \langle\Omega_{(JM)}\Omega_{(J'M')}^{*}\rangle
     =\delta_{JJ'}\delta_{MM'}C_{J}^{\Omega\Omega},
\label{eqn:potentialpowerspectra}
\end{equation}
where $\varphi_{(JM)}$ and $\Omega_{(JM)}$ are spherical-harmonic
coefficients for $\varphi(\hatn)$ and $\Omega(\hatn)$,
respectively.  The deflection field can be expanded in terms of
vector spherical waves by,
\begin{equation}
     \Delta_{a}\left(\hatn\right) = \sum_{JM}
     \sqrt{J\left(J+1\right)} \left\{ \varphi_{(JM)}Y_{(JM)a}^{E}
     \left(\hatn\right) + \Omega_{(JM)}Y_{(JM)a}^{B}
     \left(\hatn\right)\right\}.
\label{eq:deflection-multipole} 
\end{equation}

\subsection{Scalar metric perturbation}

Suppose we have a single TAM wave for a density perturbation.
This is described by a metric perturbation,
\begin{equation}
     h_{ab}(\eta',(\eta_0-\eta')\hatn) = 4 \frac{9}{10} T^{\rm
     sca}(k) \frac{D_1(\eta')}{a(\eta')} \Phi^{k,p}_{(JM)}
     \Psi^k_{(JM)}((\eta_0-\eta')\hatn) g_{ab},
\label{eqn:metricperturbation}
\end{equation}
where $D_1(\eta)$ is a linear-theory growth factor, $a(\eta)$ is
the scale factor, and $\Phi^{k,p}_{(JM)}$ is the primordial
amplitude of the Newtonian potential for wavenumber $k$ and
total angular momentum $JM$.  We then have
\begin{align}
     -\frac{1}{2} \left(\eta_{0}-\eta'\right) \hat n^b
     \hat n^c \nabla_a h_{bc} & = -2
     \left(\eta'-\eta\right) \frac{9}{10} T^{\rm sca}(k)
     \frac{D_{1} \left(\eta'\right)}
     {a\left(\eta'\right)} \left[\nabla_a\Psi_{(JM)}^{k}
     \left(\left(\eta_{0}-\eta'\right)
     \hatn\right)\right] \nonumber  \\ 
      & = 2i\left(\eta' -\eta\right) \frac{9}{10} T^{\rm
      sca}\left(k\right) \frac{D_{1} \left(\eta'\right)}
      {a\left(\eta'\right)} k \Psi_{(JM)a}^{k,L}
      \left(\left(\eta_{0}-\eta'\right) \hatn\right).
\end{align}
Using Eq.~(\ref{eqn:vectorprojections}), which provides the
projection of $\Psi^{k,L}_{(JM)}(\bfx)$ onto the plane normal to $\hat
n^a$, we find that this single $kJM$ mode of the scalar field
gives rise to a deflection field,
\begin{equation}
     \Delta_{a}^{\rm sca} \left(\hatn\right) = \sqrt{J(J+1)}
     \Phi^{k,p}_{(JM)} F_J^{\rm sca}(k) Y_{(JM)a}^{E}(\hatn),
\label{eqn:scalenstrans}
\end{equation}
with
\begin{equation}
     F_J^{\rm sca}(k) = \frac{9}{5} T^{\rm sca} \left(k\right)
     \int_{\eta_{0}}^{\eta} \, d\eta'\,
     \frac{\eta'-\eta}{\left(\eta_{0}-\eta\right)
     \left(\eta_{0} - \eta'\right)}
     \frac{D_{1}\left(\eta'\right)}
     {a\left(\eta'\right)}j_{J}\left(k\left(\eta_{0} -
     \eta'\right)\right).
\end{equation}
We thus see that a given TAM wave of $JM$ gives rise only to
spherical harmonics in the deflection field of the same $JM$.
The absence of a curl ($B$) mode is a consequence of the fact
that the longitudinal TAM wave $\Psi^L_{(JM)a}(\bfx)$ has no projection
onto $Y^B_{(JM)}(\hatn)$ [cf.~Eq.~(\ref{eqn:vectorprojections})].  We can
equivalently conclude that this particular $kJM$ TAM mode of the
potential $\Phi$ gives rise to a spherical-harmonic coefficient
$a_{(JM)}^E(k) = \Phi^{k,p}_{(JM)} F_J^{\rm sca}(k)$.  The E
mode deflection-angle power spectrum from the complete random
field is then given by summing,
\begin{equation}
     C_J^{EE} = J(J+1) C_J^{\varphi\varphi} \frac{2}{\pi} \int
     \, k^2 \, dk\, P_\Phi(k)
     \left| F_J(k) \right|^2,
\label{eqn:Epowerspectrum}
\end{equation}
over all $k$ modes with this $JM$, in agreement with results
obtained by summing over Fourier waves, rather than TAM waves.

\subsection{Tensor Metric Perturbations}

The TAM formalism will have more power, however, for tensor
metric perturbations.  So consider now a TAM wave,
\begin{equation}
     h_{ab}(\eta',(\eta_0-\eta')\hatn) = h^{k,X}_{(JM)} T(k,\eta')
     \Psi^{k,X}_{(JM)ab} \left( (\eta_0-\eta')\hatn \right),
\label{eqn:GWpert}
\end{equation}
of a transverse-traceless metric perturbation.  Here, we will
take $X$ to be either $E$ or $B$ (although we could have
alternatively considered $\lambda=\pm2$ modes in the helicity
basis), $T(k,\eta')$ gives the time evolution of modes of
wavenumber $k$, and $h^{k,X}_{(JM)}$ is the primordial amplitude
of the mode.  From Eq.~(\ref{eqn:deflection-definition}), we
will need to calculate the tangential projections of $\hat n^a
\Psi^{k,X}_{(JM)ab}(\bfx)$ and $\hat n^a \hat n^c \nabla_b
\Psi^{k,X}_{(JM)ac}(\bfx)$.  Using the transformation in
Eq.~(\ref{eqn:tensortransformations}) between the OAM and
$L/VE/VB/TE/TB$ bases and the relations,
Eq.~(\ref{eqn:tensorprojections}), between the tensor TAM waves
and the tensor spherical harmonics, one can show that
\begin{equation}
     \hat{n}^a \Psi^l_{(JM)ab}(\bfx) = \begin{cases}
     \sqrt{\frac{J-1}{2J-1}} j_{J-2}(kr) Y^{J-1}_{(JM)b}(\hatn), & \qquad l=J-2,\\
     \sqrt{\frac{J-1}{2\left(2J+1\right)}} j_{J-1}(kr) Y^{J}_{(JM)b}(\hatn), &
     \qquad l=J-1,\\
     j_J(kr) \left(
     -\sqrt{\frac{\left(J+1\right)\left(2J+3\right)}{6\left(2J-1\right)
     \left(2J+1\right)}} Y^{J-1}_{(JM)b}(\hatn) +
     \sqrt{\frac{J\left(2J-1\right)}{6\left(2J+1\right)
     \left(2J+3\right)}} Y^{J+1}_{(JM)b}(\hatn) \right), & \qquad l=J,\\
     -\sqrt{\frac{J+2}{2\left(2J+1\right)}} j_{J+1}(kr)
     Y^{J}_{(JM)b}(\hatn), &
     \qquad l=J+1,\\
     -\sqrt{\frac{J+2}{2J+3}} j_{J+2}(kr) Y^{J+1}_{(JM)b}(\hatn), &
     \qquad l=J+2.\end{cases}  
\end{equation}
It then follows that,
\begin{align}
     \hat n^a \Psi^B_{(JM)ab}(\bfx) & = \sqrt{ \frac{
     \left( J-1\right) \left(J+2\right)} {2(2J+1)^2}}
     \left(j_{J-1}(kr) + j_{J+1}(kr) \right) Y^J_{(JM)b}(\hatn) \nonumber \\
      & = -i \sqrt{\frac{\left(J-1\right) \left(J+2\right)}{2}} \frac{j_{J}(kr)
      }{k r} {Y}^{B}_{(JM)b}(\hatn),
\end{align}
which lies entirely in the plane normal to $\hatn$, and
\begin{align}
     \hat n^a \Psi^{E}_{(JM)ab}(\bfx) & = \sqrt{\frac{\left(J-1\right)
     \left(J+1\right)\left(J+2\right)} {2\left(2J-1\right)^{2} (2J+1)}}
     \left(j_{J-2}(kr)+j_{J}(kr)\right) Y^{J-1}_{(JM)b}(\hatn) \nn\\&-
     \sqrt{\frac{J\left(J-1\right) \left(J+2\right)}{2
     \left(2J+3\right)^{2} (2J+1)}} \left(j_{J}(kr)+j_{J+2}(kr)\right)
     {Y}^{J+1}_{(JM)b}(\hatn) \nonumber \\
      & =\sqrt{\frac{\left(J-1\right)
      \left(J+1\right)\left(J+2\right)}{2(2J+1)}}\frac{j_{J-1}\left(kr\right)
      }{kr} {Y}^{J-1}_{(JM)b}(\hatn) -
      \sqrt{\frac{J\left(J-1\right) \left(J+2\right)}{2(2J+1)}}
      \frac{j_{J+1}\left(kr\right)} {kr}
      {Y}^{J+1}_{(JM)b}(\hatn).
\label{eqn:ndotE}
\end{align}
This vector has components both in the normal direction and in
the tangent space.  The projection onto the tangent space is
\begin{equation}
     \Pi_{bc} n^a \Psi^{E}_{(JM)ac}(\bfx) =
     - \sqrt{\frac{\left(J-1\right) \left(J+2\right)}{2}} \frac{1}{kr}
     \left[j_{J}'\left(kr\right) +\frac{j_{J}
     \left(kr\right)}{k r}\right] {Y}^{E}_{(JM)b}(\hatn).
\end{equation}
The first term in Eq.~(\ref{eqn:deflection-definition})
contributes
\begin{equation}
     \Delta_{a}^{GW,X(1)} \left(\hatn\right) = -  h^{k,X}_{(JM)}
     \sqrt{\frac{\left(J-1\right) \left(J+2\right)}{2}}
     Y_{(JM)a}^{X} \left(\hatn\right)
     \int_{\eta_{0}}^{\eta}d\eta'
     \frac{T\left(k,\eta'\right)}
     {\left(\eta_{0}-\eta\right) k \left(\eta_{0}-\eta'
     \right)} f_{J}^X \left(k\left(\eta_{0}-\eta'\right)\right),
\end{equation}
where the radial functions are
\begin{equation}
     f_{J}^X\left(kr\right)= \begin{cases}
     i j_{J}\left(kr\right), & \qquad\mathrm{for}\, X=B,\\
     j_{J}'\left(kr\right) +\frac{ j_{J}\left(kr\right)}
     {kr}, & \qquad\mathrm{for}\, X=E.
\label{eqn:radialfunctions}
\end{cases}
\end{equation}

We now turn to the second term in
Eq.~(\ref{eqn:deflection-definition}), that proportional to
$\hat{n}_{c}\hat{n}_{d}\nabla_{e}h_{cd}$.   Using
$\nabla_{a}\hat{n}_{b}=\left(g_{ab}-\hat{n}_{a}\hat{n}_{b}\right)/r
= \Pi_{ab}/r$, we can write
\begin{align}
   \hat{n}_{c} \hat{n}_{d} \nabla_{e} h_{cd} & =\nabla_{e}
   \left(\hat{n}_{c} \hat{n}_{d} h_{cd} \right) -
   \left(\nabla_{e} \hat{n}_{c} \right) \hat{n}_{d}h_{cd} -
   \hat{n}_{c} \left(\nabla_{e} \hat{n}_{d}\right)h_{cd}
   \nonumber \\
    & = \nabla_{e} \left(\hat{n}_{c}\hat{n}_{d} h_{cd} \right)
    - 2\left(\nabla_{e} \hat{n}_{c}\right) \hat{n}_{d}h_{cd}
    \nonumber \\
    & =\nabla_{e} \left(\hat{n}_{c}\hat{n}_{d} h_{cd} \right)
    - \frac{2}{\left(\eta_{0}-\eta'\right)}\Pi_{ec} \hat{n}_{c}
    h_{cd}.
\label{eqn:projectiontensor}
\end{align}
Given that $\Pi_{ae} \Pi^e{}_c=\Pi_{ac}$, we see that the
second term here is similar to what we calculated before.  It
thus contributes,
\begin{equation}
     \Delta_{a}^{GW,X(2)} \left(\hatn\right) =- h^{k,X}_{(JM)}
     \sqrt{\frac{\left(J-1\right) \left(J+2\right)}{2}}
     Y_{(JM)a}^{X} \left(\hatn\right)
     \int_{\eta_{0}}^{\eta}d\eta'
     \frac{\left(\eta' - \eta\right)
     T\left(k,\eta' \right)}{\left( \eta_{0}-\eta\right)
     k\left(\eta_{0}-\eta'\right)^{2}}
     f_{J}^X\left(k\left(\eta_{0}-\eta'\right)\right),
\end{equation}
to the deflection field.
Now consider the first term in Eq.~(\ref{eqn:projectiontensor}).
We have already seen for the $B$ mode that $\hat{n}^a \Psi^B_{(JM)ab}(\bfx)$
is perpendicular to $\hat{n}^b$.  Thus, $\hat{n}^a \hat{n}^b
\Psi^B_{(JM)ab}(\bfx)=0$, and this term does not contribute to
the curl ($B$ mode).  Using Eq.~(\ref{eqn:ndotE}) and
\begin{equation}
     \hat{n}^a Y^{l}_{(JM)a}(\hatn) = \begin{cases}
     \sqrt{\frac{J}{2J+1}}Y_{(JM)}(\hatn), & \qquad l=J-1,\\
     0, & \qquad l=J,\\
     -\sqrt{\frac{J+1}{2J+1}}Y_{(JM)}(\hatn), & \qquad l=J+1,
\end{cases}
\end{equation}
we find 
\begin{equation}
     \hat{n}^a \hat{n}^b \Psi^{E}_{(JM)ab}(\bfx) =
      \sqrt{\frac{ (J+2)!}{2 (J-2)!}} \frac{j_{J}
      \left(kr\right)}{\left(kr\right)^{2}} Y_{JM}(\hatn).
\end{equation}
When we now take the gradient $-M_{\perp a} =r\Pi_{ab}
\nabla^b$ of this in the tangential plane, the operator acts
only on the spherical harmonic.  Using $M_{\perp a} Y_{(JM)} =
\sqrt{J(J+1)} Y^E_{(JM)a}$, we find,
\begin{equation}
     \Pi_{ab} \nabla^b \hat{n}^c \hat{n}^d \Psi^{TE}_{(JM)cd}(\bfx) = 
     -\sqrt{\frac{\left(J-1\right)\left(J+2\right)}{2}}
     \frac{J\left(J+1\right)}{k^{2}
     \left(\eta_{0}-\eta' \right)^{3}}
     j_{J}\left(k\left(\eta_{0}-\eta'\right)\right)
     {Y}^{E}_{(JM)a}(\hatn).
\end{equation}
We thus find that this term contributes
\begin{equation}
     \Delta_{a}^{GW,E(3)} \left(\hatn\right) = \frac{1}{2}
     \sqrt{\frac{ \left(J-1\right) \left(J+2\right)}{2}} J
     \left(J+1\right) 
     Y_{(JM)a}^{E}\left(\hatn\right) \int_{\eta_{0}}^{\eta}
     d\eta' \frac{\left(\eta'-\eta\right)
     T\left(k,\eta'\right)} {\left(\eta_{0}-\eta\right)
     k^{2} \left(\eta_{0}-\eta'\right)^{3}}
     j_{J}\left(k\left(\eta_{0}-\eta'\right)\right).
\end{equation}

In summary, a single B- or E-mode TAM wave contributes a
deflection field,
\begin{equation}
     \Delta_{a}^{\rm GW} \left(\hatn\right) =
     h^{k,X}_{(JM)} \sqrt{J(J+1)} F_J^{{\rm GW},X}(k)
     Y_{(JM)a}^{X}(\hatn),
\end{equation}
for $X=\{E,B\}$, with
\begin{equation}
     F_J^{{\rm GW},B}(k) =-i \sqrt{\frac{\left(J-1\right)
     \left(J+2\right)}{2 J (J+1) }}
     \frac{1}{\eta_{0}-\eta} \int_{\eta_{0}}^{\eta}
     d\eta' \frac{T\left(k,\eta'\right)}
     {k\left(\eta_{0} - \eta'\right)^2} 
		j_J\left(k\left(\eta_0-\eta'\right)\right),
\end{equation}
and
\begin{align}
     F_J^{{\rm GW},E}(k) = & -\sqrt{\frac{\left(J-1\right)
     \left(J+2\right) }{2 J (J+1) }}
		\int_{\eta_{0}}^{\eta}
     d\eta' \frac{T\left(k,\eta'\right)}{
     k\left(\eta_{0} - \eta'\right)^2}\nonumber \\
	&\times
	\left\{
		\left[j'_J\left(k\left(\eta_0-\eta'\right)\right)
			+\frac{j_J\left(k\left(\eta_0-\eta'\right)\right)}{k(\eta_0-\eta')}
		\right]
		-\frac{J(J+1)}{2}\frac{\eta'-\eta}{\eta_0-\eta}
		\frac{j_J\left(k(\eta_0-\eta')\right)}{k(\eta_0-\eta')}
	\right\}.
\end{align}
The power spectra are then obtained from
\begin{equation}
     C_J^{XX} = \frac{2}{\pi} \int \, k^2 \, dk\, \frac{P_h(k)}{2}
     \left| F_J^{{\rm GW}, X}(k) \right|^2,
\end{equation}
for $X=E,B$, by summing over all $k$ modes with this $JM$.

\section{Boltzmann equations for CMB Fluctuations}
\label{sec:cmb}

TAM waves can also be used to provide an alternative set of
Boltzmann equations to calculate CMB power spectra.  Our
discussion here is preliminary; we leave the full calculation to
future work \cite{sklee}.  There is some overlap, although
not complete, with what we discuss here and work in
Ref.~\cite{Abramo:2010za}, and also in Ref.~\cite{Aich:2010gn}.

The radiation perturbation $\Theta(\bfx,\hatq;\eta)$ is most
generally a function of position $\bfx$, the photon direction
$\hatq$, and conformal time $\eta$.  This perturbation satisfies a
Boltzmann equation, a partial differential equation in time,
space, and in photon direction $\hatq$.
In the standard treatment \cite{CMB}, one considers a single Fourier mode
$\Phi(\bfx,\eta) =\Phi_{\bf k} e^{i\bfk \cdot \bfx}$ of wavevector
$\bfk$ of the gravitational potential (or of the
gravitational-wave field).  The spatial dependence of
$\Theta(\bfx,\hatq;\eta)$ must also then be $\propto e^{i\bfk \cdot \bfx}$.  The
$\hatq$ dependence is, however, then expanded in spherical
harmonics.  Since the end result, the power spectrum $C_J$, is a rotational
invariant, one generally then chooses $\bfk \parallel \hatz$ so
that the spherical-harmonic expansion for the $\hatq$ dependence
of $\Theta_{\bf k}(\hatq;\eta)$ becomes in practice an expansion
in Legendre polynomials $P(\cos\theta_p) \propto
Y_{(J0)}(\hatq)$.

Alternatively, though, the gravitational potential can be expanded
$\Phi(\bfx,\eta)= \sum_{kJM} \Phi^k_{(JM)}(\eta)
\Psi^k_{(JM)}(\bfx)$ in terms of scalar TAM waves
$\Psi^k_{(JM)}(\bfx) = j_J(kx) Y_{(JM)}(\hatx)$ (or for tensor
perturbations, in terms of tensor TAM waves).  The most
general radiation perturbation associated with this scalar
perturbation can then be expanded in terms of states of TAM
$JM$, 
\begin{equation}
     \Theta(\bfx,\hatq;\eta) = \sum_{k,JM,ll'}
     \Theta^{k,JM}_{ll'} (\eta) \Xi^{k,JM}_{ll'}(\bfx,\hatq),
\label{eqn:radiationpert}
\end{equation}
where the total-angular-momentum eigenfunctions (which are also
eigenfunctions, of quantum numbers $l$ and $l'$, of $\bfx$ and
$\hatq$ angular momentum, respectively) are
\begin{equation}
     \Xi_{ll'}^{k,JM}(\bfx,\hatq) = \sum_{mm'} \VEV{ lml'm'|JM} j_l(kx)
     Y_{(lm)}(\hatx) Y_{(l'm')}(\hatq).
\end{equation}

It now follows that the angular dependence of the
observed radiation from a spherical wave with quantum numbers $kJM$ will be
proportional to $Y_{(JM)}(\hatq)$.  We take the observer to be
at the origin.  We then note that the radial eigenfunctions
$j_l(kr)$ all vanish at the origin unless $l=0$.  Thus,
\begin{align}
     \Theta(\hatx=0,\hatq;\eta) &= \sum_{ll'}
     \Theta^{k,JM}_{ll'} 
     \Xi^{k,JM}_{ll'} (\bfx=0,\hatq)
     = \Theta^{k,JM}_{0J}(\eta)
     \Xi^{k,JM}_{0J}(\bfx=0,\hatq) \nn \\
     &= \Theta^{k,JM}_{0J} \VEV{00JM|JM} Y_{(00)}(\hatx) j_0(0)
     Y_{(JM)}(\hatq).
\end{align}

In the TAM approach, therefore, calculation of the CMB
temperature fluctuation boils down to calculation of
$\Theta^{k,JM}_{0J}(\eta)$. The Boltzmann equation for this
particular coefficient, however, will be coupled to those for
all $\Theta^{k,JM}_{ll'}$.  We thus trade the infinite tower of
equations for the $l=0,1,2,\ldots$ coefficients $\Theta_l(k)$
for each wavenumber $k$ for an infinite tower
$l,l'=0,1,2,\ldots$ for the coefficients $\Theta^{k,JM}_{ll'}$
for a particular $J$.  The advantage,
though, is that each TAM wave of $JM$ contributes only to
$C_J$.  Thus, the power spectrum $C_J$ can be evaluated for a
single $J$.  There may also be conceptual advantages to this
approach, even if there are no immediate numerical advantages.

\section{CONCLUSIONS}
\label{sec:conclusions}

In this paper we have obtained complete sets of basis functions,
specified by their total angular momentum $JM$, for scalar,
vector, and tensor fields on ${\mathbb R}^3$.  We have written
three such sets of basis functions, one in terms of
orbital-angular-momentum states, one in terms of an $L/E/B$ or
$L/VE/VB/TE/TB$ decomposition of the vector and tensor
fields, and a third in terms of states of definite helicity.
In the process, we have also shown
how all five components of a rank-2 traceless symmetric tensor field,
including the transverse-traceless components, can be written in
terms of derivative operators acting on scalar fields, a result
that may be useful for basis functions beyond those, based on
spherical coordinates, that we have derived here.

We have shown how the projections of these three-dimensional
vector and tensor fields onto the two-sphere yield the familiar
$E/B$ vector and tensor spherical harmonics.  We found that an
$E$ mode on the two-sphere may arise from either a longitudinal
vector or tensor mode or from $E$ mode vector or tensor TAM
waves.  Conversely a $B$ mode on the two-sphere is seen to arise
from a projection of a vector or tensor $B$ mode.  We also
generalized the two usual $E/B$ tensor spherical harmonics to
account for the three other possible polarizations of a
traceless three-dimensional tensor field.  We showed how the
five TAM waves project onto these five tensor spherical
harmonics.

A realization of a random scalar, vector, or tensor field is
usually described as a collection of plane waves with amplitudes
selected from some distribution.  We have shown, however, that a
random field can also be realized as a collection of TAM waves,
and we have shown how the power spectra for these TAM-wave
amplitudes are related to the power spectra for the more
familiar plane-wave amplitudes.  The advantage of TAM waves over
the simpler but more naive outer product of the tensor spherical
harmonics with radial wave functions is that such basis
functions are not necessarily eigenfunctions of the Laplacian.
They therefore do not follow a simple evolution during the
linear regime at late times, and they are not normal modes
during inflation.

The utility of TAM waves in cosmology is apparent given that
most observations are performed on a spherical sky.  Many
calculations of cosmological observables, which are usually
performed by considering the projection of a single Fourier mode
onto a spherical sky, can be performed alternatively by
considering a single TAM wave.  The angular dependence of any
observable on a TAM wave of total angular momentum $JM$ must
then be a scalar, vector, or tensor harmonic (depending on the
observable) of that same $JM$.  We showed, as one example, how
the calculation of power spectra for the lensing-deflection
field for gravitational lensing by
density perturbations and gravitational waves is carried out in
the TAM formalism, and we made preliminary remarks on the
possible utility of the TAM formalism in numerical evaluation of
CMB power spectra.  The full power of the TAM formalism will be
manifest most clearly, though, in the calculation of
higher-order correlations (e.g., angular bispectra) in models
with non-Gaussianity, particularly those involving vector and/or
tensor fields.  The basic idea here is that the Wigner-Eckart
theorem guarantees that angular correlations of three TAM waves
must be proportional to a Clebsch-Gordan coefficient, along with
some prefactor that will depend on the tensorial nature of the
waves.  This will be presented in Ref.~\cite{inprogress}.

The development of the TAM-wave formalism requires considerable
technical detail.  However, once completed, understood, and
mastered, it may facilitate the calculation of many cosmological
observables.

\begin{acknowledgments}
We thank Samuel Lee for useful comments.
LD acknowledges the support of the Roland Research Fund.  This
work was supported by DoE SC-0008108 and NASA NNX12AE86G.
\end{acknowledgments}

\appendix

\section{Divergence of the vector harmonics}
\label{appendix:vectordivergence}
\subsection{Gradient of scalar TAM waves}
\label{appendix:scalargradient}
First, we calculate the gradient,
\be
\nabla_a \Psi^{k}_{\JM} (\bm{x})
=
\nabla_a j_{J}(kr) Y_{\JM}(\hatn),
\ee
of scalar TAM waves.
Our derivation is based on the Fourier transform.
From Rayleigh's formula [Eq.~(\ref{eq:Rayleigh})], we find 
the Fourier transform of scalar harmonics $j_J(kr)Y_\JM(\hatn)$ as
\be
\int d^3 x \,
j_{J}(kr) Y_{\JM}(\hatn)
e^{-i\bm{q}\cdot\bm{x}}
=
2\pi^2 (-i)^J
\frac{\delta_D(q-k)}{q^2} 
Y_{\JM}(\hat{\bm{q}}),
\ee
and re-write the gradient as an inverse Fourier transform:
\ba
\nabla_a \Psi_{\JM}^k(\bm{x})
&=&
\nabla_a
\int
\frac{d^3q}{(2\pi^3)}
\left[
2\pi^2 (-i)^J
\frac{\delta_D(q-k)}{q^2} 
Y_{\JM}(\hat{\bm{q}})
\right]
e^{i\bm{q}\cdot\bm{x}}
\vs
&=&
\sum_{lm}
i^{l-J+1}
\left[
\int q^2 dq 
\frac{q}{q^2}
j_{l}(qr)
\delta_D(q-k)
\right]
\left[
\int d^2\hat{\bm{q}} \,
\qhat_a Y_{\JM}(\hat{\bm{q}})
Y_\lm^*(\hat{\bm{q}})
\right]
Y_\lm(\hatn).
\ea
The radial integral trivially reads
\be
\int q^2 dq 
\frac{q}{q^2}
j_{l}(qr)
\delta_D(q-k) = k j_l(kr),
\ee
and the angular integral can be written as a Gaunt integral
by decomposing
\be
\hat{n}_a = 
\sqrt{\frac{4\pi}{3}}
\sum_{\mbar} Y_{(1\mbar)}(\hatn)(-1)^\mbar {e}_a^{-\mbar}.
\ee
Then,
\ba
\int d^2\hat{\bm{q}}\,
\qhat_a 
Y_{\JM}(\hat{\bm{q}})
Y_\lm^*(\hat{\bm{q}})
&=&
\sqrt{\frac{4\pi}{3}}
\sum_\mbar(-1)^\mbar {e}_a^{-\mbar}
\int 
d^2\qhat
Y_{(1\mbar)}(\qhat)
Y_{\JM}(\hat{\bm{q}})
Y_\lm^*(\hat{\bm{q}})
\vs
&=&
\sqrt{\frac{4\pi}{3}}
\sum_\mbar(-1)^{\mbar+m} {e}_a^{-\mbar}
\sqrt{\frac{3(2J+1)(2l+1)}{4\pi}}
\left(\begin{array}{ccc}
1      & J  & l\\
0      & 0  & 0
\end{array}\right)
\left(\begin{array}{ccc}
1      & J  & l\\
\mbar  & M  & -m
\end{array}\right)
\vs
&=&
\sqrt{\frac{4\pi}{3}}
\sum_\mbar(-1)^{M} {e}_a^{-\mbar}
\sqrt{\frac{3(2J+1)(2l+1)}{4\pi}}
\left(\begin{array}{ccc}
1      & J  & l\\
0      & 0  & 0
\end{array}\right)
\left(\begin{array}{ccc}
1    &  l  & J \\
-\mbar & m & -M 
\end{array}\right)
.
\ea
Here,
\ba
\left(\begin{array}{ccc}
l_1      & l_2  & l_3\\
m_1      & m_2  & -m_3
\end{array}\right)
\equiv
\frac{1}{\sqrt{2l_3+1}}(-1)^{l_1-l_2+m_3}
\left<
l_1m_1l_2m_2|l_3m_3
\right>,
\label{eq:Wigner3j}
\ea
is the Wigner-3$j$ symbol.
Combining all, we find the gradient of the scalar TAM wave to be
\ba
\nabla_a \Psi_{\JM}^k(\bm{x})
&=&
k\sum_{lm}
i^{l-J+1}
j_l(kr)
\sum_{\mbar}(-1)^{M} {e}_a^{-\mbar}
\sqrt{(2J+1)(2l+1)}
\left(\begin{array}{ccc}
l      & 1  & J\\
0      & 0  & 0
\end{array}\right)
\left(\begin{array}{ccc}
1    &  l  & J \\
-\mbar & m & -M 
\end{array}\right)
Y_{(lm)}(\hatn)
\vs
&=&
k
\sum_{l}
i^{l-J+1}
j_l(kr)
\sqrt{\frac{2l+1}{2J+1}}
\left<l010|J0\right>
\sum_{m\mbar} {e}_a^{-\mbar}
\left<1-\mbar lm|JM\right>
Y_{(lm)}(\hatn)
\vs
&=&k
\left[
\sqrt{\frac{J}{2J+1}} 
\Psi_{\JM a}^{J-1}(\bm{x})
+
\sqrt{\frac{J+1}{2J+1}} 
\Psi_{\JM a}^{J+1}(\bm{x})
\right],
\label{eqn:divofscalar}
\ea
where we have used that the relevant Clebsch-Gordan coefficient
is non-zero only for 
\be
\left<l010|J0\right>=
\left\{
\begin{array}{ll}
\sqrt{\frac{l+1}{2l+1}}, & l = J-1, \\
-\sqrt{\frac{l}{2l+1}}, & l = J+1.
\end{array}
\right.
\ee
\subsection{Divergence of vector TAM waves}

We can now derive the divergence, given in Eq.~(\ref{eq:div}), 
of the vector TAM waves.  We start with
\ba
r\nabla^a \Psi^{l,k}_{\JM a}(\bm{x})
&=&
r\nabla^a 
\left[
j_l(kr)
\sum_{m,\mbar}
\left<
1\mbar lm|JM
\right> Y_\lm(\hatn) e_a^\mbar
\right]
\vs
&=&
\sum_{m,\mbar}
\left<
1\mbar lm|JM
\right> 
r{e}_a^\mbar \nabla^a 
j_l(kr)
Y_\lm(\hatn),
\ea
and then from Eq.~(\ref{eqn:divofscalar}),
we have 
\ba
r\nabla^a \Psi^{l,k}_{\JM a}(\bm{x})
&=&
kr
\sum_{m,\mbar}
\left<
1\mbar lm|JM
\right> 
{e}_a^\mbar 
\left[
\sqrt{\frac{l}{2l+1}} \Psi_{(lm) a}^{l-1}(\bm{x})
+
\sqrt{\frac{l+1}{2l+1}} \Psi_{(lm) a}^{l+1}(\bm{x})
\right].
\ea
The sum can be simplified as
\ba
\sum_{m,\mbar} & &
\left<
1\mbar lm|JM
\right> 
{e}_a^\mbar 
\Psi_{(lm) a}^{l\pm 1}(\bm{x})
=
j_{l\pm1}(kr)
\sum_{m,\mbar}
\left<
1\mbar lm|JM
\right> 
{e}_a^\mbar 
\sum_{m',\mbar'}\langle 1,l\pm1;\mbar'm'|lm\rangle  Y_{(l\pm1,m')}(\hatn){e}_a^{\mbar'}
\vs
& = &
j_{l\pm1}(kr)
\sum_{m,\mbar}
\sum_{m',\mbar'}
\left<
1\mbar lm|JM
\right> 
\langle 1,l\pm1;\mbar'm'|lm\rangle  Y_{(l\pm1,m')}(\hatn)
(-1)^\mbar\delta_{\mbar,-\mbar'}
\vs
&= &
j_{l\pm1}(kr)
\sum_{m,\mbar}
\sum_{m'}
(-1)^\mbar
\left<
1\mbar lm|JM
\right> 
\langle 1,l\pm1;-\mbar m'|lm\rangle  
Y_{(l\pm1,m')}(\hatn)
\vs
&=&
j_{l\pm1}(kr)
\sum_{m,\mbar}
\sum_{m'}
\sqrt{(2J+1)(2l+1)}(-1)^{\mbar+M+m-1}
\left(\begin{array}{ccc}
1    &  l  & J \\
\mbar & m & -M 
\end{array}\right)
\left(\begin{array}{ccc}
1    &  l\pm1  & l \\
-\mbar & m' & -m
\end{array}\right)
Y_{(l\pm1,m')}(\hatn)
\vs
&= &
j_{l\pm1}(kr)
\sum_{m,\mbar}
\sum_{m'}
\sqrt{(2J+1)(2l+1)}(-1)^{\mbar-1+M+m}
\left(\begin{array}{ccc}
1    &  l  & J \\
\mbar & m & -M 
\end{array}\right)
\left(\begin{array}{ccc}
1    &  l   &l\pm1  \\
\mbar & m   & -m' 
\end{array}\right)
Y_{(l\pm1,m')}(\hatn)
\vs
&= &
-
j_{l\pm1}(kr)
\sum_{m'}
\sqrt{2J+1}
\sqrt{2l+1}
\left[
\sum_{m,\mbar}
\left(\begin{array}{ccc}
1    &  l  & J \\
\mbar & m & -M 
\end{array}\right)
\left(\begin{array}{ccc}
1    &  l   &l\pm1  \\
\mbar & m   & -m' 
\end{array}\right)
\right]
Y_{(l\pm1,m')}(\hatn)
\vs
&=&
-
j_{l\pm1}(kr)
\sum_{m'}
\sqrt{2J+1}
\sqrt{2l+1}
\frac{\delta_{J,l\pm1}\delta_{Mm'}}{2J+1}
Y_{(l\pm1,m')}(\hatn)
=
-
j_{l\pm1}(kr)
\sqrt{\frac{2l+1}{2J+1}}
\delta_{J,l\pm1}
Y_{(l\pm1,M)}(\hatn),
\ea
from which follows Eq.~(\ref{eq:div}).

\section{Divergence of tensor harmonics}
\label{appendix:tensordivergence}

In this Appendix we return to the use of our usual index notation
for vectors and tensors so that vectors and tensors can be
distinguished by the number of indices.  The divergence of
the tensor TAM waves is,
\be
     \frac{1}{k}\nabla^a \Psi^l_{(JM)ab}(\bfx) = \sum_{\tilde{m}m}
     \langle 2\tilde{m}lm|JM \rangle \frac{1}{k} \left(\nabla^a
     j_{l}(kr) Y_{(lm)}(\hatx) \right) {t}^{\tilde{m}}_{ab} =
     \sum_{\tilde{m}m} \langle2\tilde{m}lm|JM \rangle
     \Psi^{L\,a}_{(lm)}(\bfx) {t}^{\tilde{m}}_{ab}.
\ee
From Eq.~(\ref{eqn:divofscalar}), we calculate
\begin{align}
     \Psi^{l'\,a}_{(lm)} t^{\tilde{m}}_{ab} &= \sum_{m'\bar{m}}
     \sum_{\bar{m}_{1}\bar{m}_{2}} \langle1\bar{m}l'm'|lm\rangle
     \Psi_{(l'm')}(\bfx)
     \langle1\bar{m}_{1}1\bar{m}_{2}|2 \tilde{m}\rangle
     e^{\bar{m}\,a} \left(e^{\bar{m}_{1}}_a
     e^{\bar{m}_{2}}_b\right)\nn\\
     & = \sum_{\bar{m}_{2}} \sum_{m'\bar{m}} (-1)^{\bar{m}}
     \langle1,-\bar{m},1\bar{m}_{2} | 2\tilde{m}\rangle \langle
     1\bar{m}l'm' | lm \rangle \Psi_{(l'm')}(\bfx) e^{\bar{m}_{2}}_b.
\end{align}
We first work out the sums over $m$, $\bar{m}$, and $\tilde{m}$. We
trade Clebsch-Gordan coefficients for Wigner-3$j$ symbols,
\begin{align}
     &\sum_{m\bar{m}\tilde{m}}(-1)^{\bar{m}}\langle2
     \tilde{m}lm|JM\rangle\langle1,-\bar{m},1\bar{m}_{2}|2
     \tilde{m}\rangle\langle1\bar{m}l'm'|lm\rangle\nn\\
     &=\sum_{m\bar{m}\tilde{m}}(-1)^{1+l+l'+M+\bar{m}+m+\tilde{m}}
     \sqrt{5\left(2l+1\right)\left(2J+1\right)}\left(\begin{array}{ccc}
     2 & l & J\\
     \tilde{m} & m & -M\end{array}\right)\left(\begin{array}{ccc}
     1 & 1 & 2\\
     -\bar{m} & \bar{m}_{2} &
     -\tilde{m}\end{array}\right)\left(\begin{array}{ccc}
     1 & l' & l\\
     \bar{m} & m' & -m\end{array}\right)\nn\\
     &=(-1)^{M+J+1}\sqrt{5\left(2l+1\right)
     \left(2J+1\right)} \sum_{m\bar{m}\tilde{m}}
     (-1)^{l+m+1+\bar{m}+2+\tilde{m}}\left(\begin{array}{ccc}
     1 & 2 & 1\\
     \bar{m}_{2} & -\tilde{m} &
     -\bar{m}\end{array}\right)\left(\begin{array}{ccc}
     l & l' & 1\\
     -m & m' & \bar{m}\end{array}\right)\left(\begin{array}{ccc}
     l & 2 & J\\
     m & \tilde{m} & -M\end{array}\right)\nn\\
     &=(-1)^{M+J+1} \sqrt{5\left(2l+1\right)\left(2J+1\right)}
     \left\{ \begin{array}{ccc}
     1 & l' & J\\
     l & 2 & 1\end{array}\right\} \left(\begin{array}{ccc}
     1 & l' & J\\
     \bar{m}_{2} & m' & -M\end{array}\right),
\end{align}
where we have used the definition,
\begin{align}
&\left\{ \begin{array}{ccc}
     L_{1} & L_{2} & L_{3}\\
     l_{1} & l_{2} & l_{3}\end{array}\right\} \left(\begin{array}{ccc}
     L_{1} & L_{2} & L_{3}\\
     M_{1} & M_{2} & M_{3}\end{array}\right)\nn\\
     &\qquad \equiv\sum_{m_{1}m_{2}m_{3}}(-1)^{l_{1}+l_{2}+l_{3}+m_{1}+m_{2}+m_{3}}
     \left(\begin{array}{ccc}
     L_{1} & l_{2} & l_{3}\\
     M_{1} & m_{2} & -m_{3}\end{array}\right)\left(\begin{array}{ccc}
     l_{1} & L_{2} & l_{3}\\
     -m_{1} & M_{2} & m_{3}\end{array}\right)\left(\begin{array}{ccc}
     l_{1} & l_{2} & L_{3}\\
     m_{1} & -m_{2} & M_{3}\end{array}\right),
\label{eq:Wigner6j}
\end{align}
for the Wigner-6$j$ symbol.  We are then left with
\begin{align}
     \sum_{\bar{m}_{2}m'}(-1)^{M+J+1} &
     \sqrt{5\left(2l+1\right)\left(2J+1\right)}\left\{ \begin{array}{ccc}
     1 & l' & J\\
     l & 2 & 1\end{array}\right\} \left(\begin{array}{ccc}
     1 & l' & J\\
     \bar{m}_{2} & m' & -M\end{array}\right) \Psi_{(l'm')}(\bfx)
     e^{\bar{m}_{2}}_a \nn\\
     &=-(-1)^{J+l'+1}\sqrt{5\left(2l+1\right)}
     \left\{ \begin{array}{ccc}
     1 & l' & J\\
     l & 2 & 1\end{array}\right\}
     \sum_{\bar{m}_{2}m'}\langle1\bar{m}_{2}l'm'|JM\rangle \Psi_{(l'm')}(\bfx)
     {e}^{\bar{m}_{2}}_a \nn\\ 
     &=-(-1)^{J+l}\sqrt{5\left(2l+1\right)}
     \left\{ \begin{array}{ccc}
     1 & l' & J\\
     l & 2 & 1\end{array}\right\} \Psi^{l'}_{(JM)a}(\bfx).
\end{align}
We evaluate the 6$j$-symbols explicitly and find
\begin{align}
     \frac{1}{k}\nabla^a \Psi^l_{(JM)ab}(\bfx)= &-\sqrt{5}(-1)^{J+l}
     \left(\left\{ \begin{array}{ccc}
     1 & l-1 & J\\
     l & 2 & 1\end{array}\right\}
     \sqrt{l} \Psi^{l-1}_{(JM)b}(\bfx)+\left\{ \begin{array}{ccc}
     1 & l+1 & J\\
     l & 2 & 1\end{array}\right\} \sqrt{l+1}
     \Psi^{l+1}_{(JM)b}(\bfx)\right)\nn\\
     &=-\begin{cases}
     \sqrt{\frac{J-1}{2J-1}} \Psi^{J-1}_{(JM)b}(\bfx), & \qquad l=J-2,\\
     \sqrt{\frac{J-1}{2\left(2J+1\right)}} \Psi^{J}_{(JM)b}(\bfx), &
     \qquad l=J-1,\\
     \sqrt{\frac{\left(J+1\right)\left(2J+3\right)}{6\left(2J-1\right)
     \left(2J+1\right)}} \Psi^{J-1}_{(JM)b}(\bfx) +
     \sqrt{\frac{J\left(2J-1\right)}{6\left(2J+1\right)
     \left(2J+3\right)}} \Psi^{J+1}_{(JM)b}(\bfx), & \qquad l=J,\\
     \sqrt{\frac{J+2}{2\left(2J+1\right)}} \Psi^{J}_{(JM)b}(\bfx), &
     \qquad l=J+1,\\
     \sqrt{\frac{J+2}{2J+3}} \Psi^{J+1}_{(JM)b}(\bfx), & \qquad
     l=J+2.\end{cases}
\end{align}
Note that we automatically obtain eigenfunctions of total
angular momentum as a consequence of acting with $\nabla_a$, an
irreducible-vector operator, on a tensor eigenfunction of total
angular momentum.

\section{Irreducible tensors}
\label{appendix:irreducible} 

In this Appendix, we show that irreducible-tensor operators,
when acting on a wavefunction, conserve the total angular
momentum, even though the spin might change. 
To be more precise, let us consider the spherical harmonics
$Y_{(JM)}(\hatn)$, which are eigenfunctions of orbital angular
momentum $\mathbf{L}^2$ and $L_z$, for given $JM$; i.e.,
\be
    \mathbf{L}^{2}Y_{(JM)}=J\left(J+1\right)Y_{(JM)},\qquad
    L_{z}Y_{(JM)}=MY_{(JM)}.
\ee
Assume we have a group of irreducible-tensor operators
$\mathcal{O}^l_m$, for $m=-l,-l+1,\ldots,l-1,l$, that transform
as a representation of order $l$ under rotations. There are $2l+1$
such operators, and $\mathcal{O}^l_m Y_{(JM)}$ is a spin-$l$
object, a tensor wavefunction of higher rank. There will be
spin operators $S_a$, for $a=x,y,z$, that act on such spin-$l$
objects. The total angular momentum  $J_a=L_a+S_a$ is then defined
as the sum of the orbital angular momentum and the spin. We
would like to prove that
\be
\label{eq:proof-objective}
     \mathbf{J}^{2}\mathcal{O}_{m}^{l}Y_{(JM)}  =
     J\left(J+1\right)\mathcal{O}_{m}^{l}Y_{(JM)},\qquad 
     J_{z}\mathcal{O}_{m}^{l}Y_{(JM)}=M\mathcal{O}_{m}^{l}Y_{(JM)}.
\ee

Consider a rotation $\mathcal{R}=e^{i\Theta^a J_a}$ acting on
the Hilbert space of spin-$l$ wavefunctions, where $\Theta^a$
parametrize rotation angles. The orbital angular momentum $L_a$
generates rotations of configuration space, and the spin $S_a$
generates rotations of the internal tensor space; i.e. mixing of
the tensor components. Using the fact that $L_a$ and $S_a$
commute, we have
\be
     e^{i\Theta^{a}J_{a}}\mathcal{O}_{m}^{l} Y_{(JM)}=
     e^{i\Theta^{a}L_{a}} \left(e^{i\Theta^b
     S_b}\mathcal{O}_{m}^{l}\right) e^{-i\Theta^c
     L_c}e^{i\Theta^d L_d}Y_{(JM)}.
\ee
First we note that spin operators rotate $\mathcal{O}_{m}^{l}$
contravariantly by acting on the tensor index $m$ (if the tensor
basis transforms covariantly),
\be
e^{i\Theta^{a}S_{a}}\mathcal{O}_{m}^{l}=\sum_{m_{1}}\mathcal{O}_{m_{1}}^{l}\mathcal{D}_{mm_{1}}^{l*}\left(\mathcal{R}\right),
\label{eqn:wignerrotation}
\ee
where we have introduced the Wigner $\mathcal{D}$ rotation
matrices $\mathcal{D}^l_{m_1 m_2}\left(\mathcal{R}\right)$. Meanwhile,
the orbital angular momentum rotates $Y_{(JM)}$ covariantly,
\be
     e^{i\Theta^{a}L_{a}}Y_{(JM)}=\sum_{M_{1}}Y_{(JM_{1})}\mathcal{D}_{M_{1}M}^{J}\left(\mathcal{R}\right).
\ee
Finally, $\mathcal{O}_{m}^{l}$, being irreducible-tensor
operators, are rotated by the orbital-angular-momentum operators
according to,
\be
     e^{i\Theta^{a}L_{a}}\mathcal{O}_{m_{1}}^{l} e^{-i\Theta^b
     L_b}=\sum_{m_{2}}\mathcal{O}_{m_{2}}^{l}\mathcal{D}_{m_{2}m_{1}}^{l}\left(\mathcal{R}\right).
\ee
From the above three equations, plus the unitarity of Wigner
$\mathcal{D}$-matrices,
\be
\sum_{m_{1}}\mathcal{D}_{mm_{1}}^{l*}\left(\mathcal{R}\right)\mathcal{D}_{m_{2}m_{1}}^{l}\left(\mathcal{R}\right)=\delta_{mm_{2}},
\ee
we find
\be
e^{i\Theta^{a}J_{a}}\mathcal{O}_{m}^{l}Y_{(JM)}=\sum_{M_{1}}\mathcal{O}_{m}^{l}Y_{(JM_{1})}\mathcal{D}_{M_{1}M}^{J}\left(\mathcal{R}\right).
\ee
The derivation holds for any rotation $\mathcal{R}$, so we
conclude that spin-$l$ objects $\mathcal{O}_{m}^{l}Y_{(JM)}$
transform as a representation of order $J$ under rotations, and
hence must be eigenfunctions of total angular momentum, as
described in Eq.(\ref{eq:proof-objective}).

The proof is easily generalized if the $Y_{(JM)}$ are
replaced by spherical harmonics of higher spin. Then the orbital
angular momentum $L_a$ is replaced by total angular momentum,
while $J_a$ will be the new total angular momentum which is
obtained by adding the additional spin carried by
irreducible-tensor operators ${O}_{m}^{l}$.

\end{document}